%% file: psl22wzw_arxive.tex
\documentclass[12pt,reqno]{article}
\usepackage{amsmath}
\usepackage{amsthm}
\usepackage{amsfonts}
\usepackage{amssymb}
\usepackage{graphics,color}
\input{xy}
\xyoption{all}

\numberwithin{equation}{section}

\newtheorem{lemma}{Lemma}
\newtheorem{conjecture}{Conjecture}

\newcommand\pl{\partial}
\def\a{\alpha}
\def\b{\beta}
\def\c{\gamma}
\def\d{\delta}
\newcommand\e{\epsilon}

\def\sfrac12{{\scriptstyle \frac12}}
\newcommand{\foh}{{\scriptstyle \frac12}}
\newcommand{\fth}{{\scriptstyle \frac32}}
\newcommand{\ffh}{{\scriptstyle \frac52}}
\def\P{{\mathcal{P}}}
\newcommand{\aP}{{\mathcal{\hat{P}}}}

\newcommand{\bartial}{{\bar{\partial}}}

\newcommand{\Real}{\mathbb{R}}

\newcommand{\Integer}{\mathbb{Z}}
\newcommand{\mat}{\begin{pmatrix}}
\newcommand{\tam}{\end{pmatrix}}

\newcommand{\mc}{\mathcal}

\newcommand{\Htyp}{\mathcal{H}_{\text{typ}}}
\newcommand{\Hatyp}{\mathcal{H}_{\text{atyp}}}

\newcommand\id{\text{id}}
\newcommand\tr{\text{tr}}
\newcommand\str{\text{str}}
\newcommand\sdet{\text{sdet}}
\newcommand\WZNW{{\text{WZNW}}}
\newcommand\g{\mathfrak{g}}

\newcommand{\AdS}{{\text{AdS}}}

\newcommand{\GL}{{\text{GL}(1|1)}}
\newcommand{\gl}{{\text{gl}(1|1)}}
\newcommand{\GLmn}{{\text{GL}(m|n)}}

\newcommand{\SLmn}{{\text{SL}(m|n)}}
\newcommand{\SLnn}{{\text{SL}(n|n)}}
\newcommand{\sll}{{\text{sl}(2)}}
\newcommand{\slto}{{\text{sl}(2|1)}}
\newcommand{\su}{{\text{su}(2)}}

\newcommand\SU{{\text{SU}(2)}}
\newcommand\PSL{{\text{PSL}(2|2)}}
\newcommand\PSLnn{{\text{PSL}(n|n)}}
\newcommand\PSU{{\text{PSU}(1,1|2)}}
\newcommand\psl{{\text{psl}(2|2)}}
\newcommand\psu{{\text{psu}(1,1|2)}}
\newcommand\apsl{{\widehat{\text{psl}}}}
\newcommand\asl{{\widehat{\text{sl}}(2)}}
\newcommand\apsu{{\widehat{\text{psu}}}}
\newcommand\asu{{\widehat{\text{su}}(2)}}
\newcommand\psin{{\psi_0^\lambda}}

\title{
\bf The WZNW model on $\mathbf{PSU(1,1|2)}$}
\author{\\[0mm] Gerhard G\"otz$^1$, Thomas Quella$^{2,3}$,
               Volker Schomerus$^{1,4}$ \\[8mm]
\small$^1$ Service de Physique Th\'eorique, CEA Saclay,\\
\small F-91191 Gif-sur-Yvette, France\\[5mm]
\small $^2$ King's College London, Department of Mathematics, \\
\small Strand, London WC2R 2LS, United Kingdom\\[5mm]
\small $^3$ KdV Institute for Mathematics, University of Amsterdam,\\
\small Plantage Muidergracht 24, 1018 TV Amsterdam, The Netherlands\\[5mm]
\small $^4$ DESY Theory Group, DESY Hamburg,\\
\small Notkestrasse 85, D--22603 Hamburg, Germany}
\date{}
\begin{document}
\begin{titlepage}      \maketitle       \thispagestyle{empty}

\begin{abstract}
According to the work of Berkovits, Vafa and Witten, the
non-linear sigma model on the supergroup $\PSU$ is the essential
building block for string theory on $\AdS_3 \times \text{S}^3
\times \text{T}^4$. Models associated with a non-vanishing value
of the RR flux can be obtained through a $\psu$ invariant marginal
deformation of the WZNW model on $\PSU$. We take this as a
motivation to present a manifestly $\psu$ covariant construction
of the model at the Wess-Zumino point, corresponding to a purely
NSNS background 3-form flux. At this point the model possesses an
enhanced $\apsu(1,1|2)$ current algebra symmetry whose
representation theory, including explicit character formulas, is
developed systematically in the first part of the paper. The space
of vertex operators and a free fermion representation for their
correlation functions is our main subject in the second part.
Contrary to a widespread claim, bosonic and fermionic fields are
necessarily coupled to each other. The interaction changes the
supersymmetry transformations, with drastic consequences
for the multiplets of localized normalizable states in the model.
It is only this fact which allows us to decompose the full state space
into multiplets of the global supersymmetry. We analyze these
decompositions systematically as a preparation for a forthcoming
study of the RR deformation.
\end{abstract}
\vspace*{-23cm} {\tt hep-th/0610070\hfill DESY 06-147\\
KCL-MTH-06-09\hfill{SPhT-T06/049}\\ZMP-HH/06-016}
\bigskip\vfill
\noindent
\phantom{wwwx}{\footnotesize e-mail: }{\footnotesize\tt
 Gerhard.Goetz@cea.fr, vschomer@mail.desy.de, tquella@science.uva.nl }

\end{titlepage}

\baselineskip=19pt
\setcounter{equation}{0}
\section{Introduction}

String theory duals to superconformal field theories in various
dimensions (see \cite{Aharony:1999ti} for a review) can be related
to 2D sigma models on supergroups and cosets (see \cite{Metsaev:1998it,
Rahmfeld:1998zn,Berkovits:1999im,Berkovits:1999zq,Berkovits:2000fe,
Roiban:2000yy} for some early references). The precise relation depends
on the particular framework, i.e.\ whether the models arise within the
Green-Schwarz formalism, the hybrid or the pure spinor approach.
These developments provide strong motivation to study world-sheet
models with supermanifolds as target spaces. This applies in
particular to 1+1 dimensional sigma models on the superconformal
group $\PSU$. In this case, the hybrid formalism developed by
Berkovits, Vafa and Witten \cite{Berkovits:1999im} furnishes a
covariant construction of string theory on $\AdS_3 \times
\text{S}^3$. The main constituent of their formulation is a
sigma model on $\PSU$.\footnote{The same model has been proposed
to describe plateau transitions in the integer quantum Hall
effect \cite{Zirnbauer:1999ua} (see also \cite{Bhaseen:1999nm,
Guruswamy:1999hi} for some further studies in this context).
Let us note that models with superalgebra symmetries arise
quite generally when systems with disorder are
treated using Efetov's supersymmetric method
\cite{Efetov:1983xg} (see also \cite{Bernard:1995as} for
a review).}
\smallskip

Type IIB string theory on $\AdS_3 \times\text{S}^3$ has one rather
peculiar feature, namely that the conditions on background fields
imposed by the string equations of motion may be solved by both RR
and/or NSNS 3-form fluxes. Hence, there exists a 2-parameter
family of $\AdS_3\times\text{S}^3$ backgrounds with an unbroken
$\PSU$ symmetry. It is well known that models with pure NSNS
background fields are easiest to deal with and indeed string
theory in $\AdS_3 \times\text{S}^3$ has been solved for such cases
using the NSR formalism \cite{Maldacena:2000hw,Maldacena:2001km},
based on earlier work on the Euclidean model
\cite{Gawedzki:1991yu, Teschner:1997ft,Teschner:1999ug}. According
to common folklore, however, incorporating RR fluxes in the NSR
formulation is conceptually difficult. This is where the hybrid
approach comes in: it essentially removes the conceptual issues,
but certainly leaves us with the hard task of solving the
non-linear sigma model on $\PSU$.
\smallskip

Though very little is known about sigma models on superspaces,
there exist a few interesting results that are particularly
relevant in our present context. Most importantly, Bershadsky et
al. \cite{Bershadsky:1999hk} have argued that quantum theories
with $\text{PSL}(N|N)$ target space are conformally invariant even
before including the familiar WZ term. Of course the latter may
then be added with any integer coefficient, preserving conformal
invariance. Such a behavior can ultimately be traced back to the
vanishing of the dual Coxeter number of $\text{PSL}(N|N)$ along
with the uniqueness of the invariant rank $3$ tensor. This
observation fits nicely with the before-mentioned parametrization
of $\AdS_3 \times\text{S}^3$ backgrounds. WZNW models on $\PSU$ at
level $k$ describe pure NSNS backgrounds with $N=k+2$ units of
NSNS flux running through the 3-sphere. Varying the coefficient of
the kinetic relative to the WZ term corresponds to adding RR flux,
see \cite{Berkovits:1999im} for a precise relation between the
parameters. Hence, the hybrid formulation offers a conceptually
very simple description of $\AdS_3 \times\text{S}^3$ backgrounds
with RR flux through marginal deformations of $\PSU$ WZNW models.
Let us note that the parameter associated with RR fluxes is
continuous in perturbative string theory since the mass of
D5-branes is suppressed by a factor $g_s$ relative to the mass of
NS5-branes.
\smallskip

Obviously, the construction of sigma models on $\PSU$ through
marginal deformation of the WZNW theory remains a very difficult
technical problem. To begin with, surprisingly little is known
even about WZNW models on supergroups. As we shall demonstrate
below, models with current superalgebra symmetries behave very
differently from their bosonic counterparts. The second obstacle
arises with the RR deformation which is still technically hard to
control since it breaks many of the local symmetries of the
underlying world-sheet model. In fact, it was shown in
\cite{Bershadsky:1999hk} that switching on the deformation reduces
the chiral symmetries of the conformal field theory from a full
$\psu$ current algebra to the chiral algebra generated by the
so-called Casimir fields which is too small a symmetry to render
the model solvable within a standard conformal field theory
analysis.\footnote{In \cite{Bershadsky:1999hk}, the misnomer
``Casimir algebra'' was used for the generic chiral symmetry of
the deformed models. This deviates from standard conventions. In
fact, except for very special cases the Casimir algebra is much
larger than the algebra of Casimir fields (see
\cite{Bouwknegt:1993wg} for a nice review).} Nevertheless, some
conformal field theory techniques, and in particular conformal
perturbation theory, do offer a promising approach to computing
certain spectra in the theory, even at generic points in the
moduli space. We shall come back to this issue in a forthcoming
paper.
\smallskip

The main focus of this work is on the $\psu$ covariant
construction of the WZNW model on the $\PSU$. We exploit and
extend the insights which have been gained recently in
\cite{Schomerus:2005bf} where the WZNW on $\GL$ has been
re-examined using a free field representation. In
comparison to the earlier solution of the $\GL$ WZNW model by
Rozansky and Saleur \cite{Rozansky:1992rx}, the new approach
linked some of the peculiar properties of the field theory to
characteristic features of super-geometry. In this geometric
context, it can be argued in particular that WZNW models on
supergroups give rise to examples of logarithmic conformal field
theories (see e.g.\
\cite{Gurarie:1993xq,Gaberdiel:1998ps,Flohr:2001zs,
Gaberdiel:2001tr} and references therein). The appearance of
logarithmic singularities had been observed repeatedly before in
the theory of disordered systems (see e.g.\
\cite{Caux:1995nm,Maassarani:1996jn,Caux:1996kq}). Another
property of the $\GL$ WZNW model that was also established in
\cite{Schomerus:2005bf} is its symmetry with respect to a special
spectral flow automorphism of the current superalgebra. We shall
encounter the same features in the $\PSU$ WZNW model, though the
derivation is a bit different due to the non-compactness of the
target space. The logarithmic singularities turn out to affect
only the sector of localized normalizable states in the theory. It
is a somewhat surprising outcome of this analysis that -- contrary
to a widespread believe, see e.g.\ \cite{Bhaseen:1999nm}
-- the WZNW model on $\PSU$ does not simply
factorize into a product of the usual bosonic subsector and a
bunch of free fermions. Such a factorization applies only to the
free field theory that is used in the construction, but receives
an interesting correction due to a non-trivial screening charge.
The latter modifies, in particular, the transformation laws of
fields in a rather non-trivial way. This fact becomes crucial for
a successful RR deformation of the theory (see below and
our forthcoming paper).
\smallskip

We have decided to separate the material of this paper
into two parts. The first contains a rather complete
discussion of the representation theoretic foundations
for both the finite dimensional Lie algebra $\psl$ and its
affine version $\apsl(2|2)_k$. Special attention is devoted
to infinite dimensional representation of $\psl$. Among the
main new results are explicit character formulas for all
irreducible representations of $\apsl(2|2)_k$ belonging
to finite dimensional representations and the infinite
dimensional discrete and principal series. The second part
then deals with the $\PSU$ WZNW model. After an
extended discussion of the action functional, we study the state
space first in the minisuperspace approximation. It is
shown that the Laplacian on $\PSU$ is non-diagonalizable,
and the structure of the Jordan blocks is discussed. In fact,
we shall provide explicit formulas for all its generalized
eigenfunctions and study their transformation law wrt.\ %
the action of $\psl$. Following this discussion, we explain how
correlators of the WZNW model on $\PSU$ can be computed starting
from correlation functions for the WZNW model on the bosonic base.
We shall also see how the non-trivial properties of the
minisuperspace theory re-emerge in the field theory, giving rise
to those features of the WZNW model we have outlined in the
previous paragraph. Finally, as an application of our main
results, we shall address the Casimir decomposition of the state
space. More precisely, we describe an algorithm that allows to
count all the states of the theory which transform in the same
representation with respect to the global symmetries. These
results shall serve as a starting point for a forthcoming analysis
of the RR deformation.
\newpage

\noindent
{\huge{\bf Part I: Representation theory} }\\[2cm]
The first part of this work is devoted to the representation
theory of both the finite dimensional Lie superalgebra $\psl$
and its affine counterpart. We shall discuss finite and
infinite dimensional representations of $\psl$ and the
corresponding modules of the $\psl$ current algebra. Some
results on the finite dimensional representations of $\psl$
are fairly standard but they are included for completeness
(see \cite{MR1378540,Germoni1998:MR1659915,Zhang:2004qx,Gotz:2005ka}
for more details and references).
We believe that our analysis of representations of the
affine algebra and their characters are new.

\section{Representation theory of $\psl$}

In this section we shall discuss the Lie superalgebra $\psl$ and its
finite and infinite dimensional representations. The latter come
in two series, namely a principal continuous and a `discrete' series.
We will not comment on the complementary series since it does
not have any physical significance in the context we are interested in.

\subsection{The Lie superalgebra $\psl$}

The Lie superalgebra $\g=\psl$ possesses six bosonic generators $K^{ab}
= - K^{ba}$ with $a,b= 1,\dots,4$. They form the Lie algebra $\text{so}(4)$
which is isomorphic to $\g^{(0)}=\sll\oplus\sll$. In addition, there are eight
fermionic generators that we denote by $S^a_\a$. They split into two
sets ($\a=1,2$) each of which transform in the vector representation
of $\text{so}(4)$ ($a=1,\dots,4$) which is the $(1/2,1/2)$ of $\sll\oplus\sll$.
The relations of $\psl$ are then given by
\begin{equation}
  \label{eq:CR}
  \begin{split}
[K^{ab},K^{cd}] &\ =\  i \left[ \d^{ac}K^{bd}-\d^{bc}K^{ad}-
                        \d^{ad}K^{bc}+\d^{bd}K^{ac}\right] \\[2mm]
[K^{ab},S^c_\c] &\ =\  i \left[ \d^{ac} S^b_\c - \d^{bc} S^a_\c
   \right] \\[2mm]
[ S^a_\a,S^b_\b ] &\ =\  \frac{i}{2}\,\e_{\a\b}\, \e^{abcd} K^{cd} \ \ .
  \end{split}
\end{equation}
Here, $\e_{\a\b}$ and $\e^{abcd}$ denote the usual completely
antisymmetric $\e$-symbols with $\epsilon_{12}=1$ and
$\epsilon^{1234}=1$, respectively. An invariant metric is
given by
\begin{align}
  \label{eq:Metric}
  \langle K^{ab},K^{cd}\rangle&\ =\ -\epsilon^{abcd}&
  \langle S_\a^a,S_\b^b\rangle&\ =\ -\epsilon_{\a\b}\,\delta^{ab}\ \ .
\end{align}
It is unique up to a scalar factor. The signs have been chosen
in view of the real form $\psu$ which will be considered
below. In order to define a root space decomposition of $\psl$ we
split the fermions $\g^{(1)}$ into two sets of four generators
$$ \g^{(1)}_+ =  {\text{span}}\{S_1^a\}\ \ \ , \ \ \
   \g^{(1)}_- =  {\text{span}}\{S_2^a\}\ \ .
$$
As indicated by the subscripts $\pm$, we shall think of the fermionic
generators $S_1^a$ as annihilation operators and of $S_2^a$ as creation
operators.
\smallskip

In our discussion below we shall also employ a second basis which
clearly exhibits the $\sll\oplus\sll$ structure of the bosonic
subalgebra. The two Cartan generators of this new basis are given
by
\begin{align}
  \label{eq:RootOne}
  K_1^0&\ =\ \frac{1}{2}\bigl[K^{12}+K^{34}\bigr]&
  K_2^0&\ =\ \frac{1}{2}\bigl[K^{12}-K^{34}\bigr]\ \ .
\end{align}
These are supplemented by the bosonic raising and lowering operators
of the form
\begin{equation}
  \begin{split} \label{KxKmn}
    K_1^\pm&\ =\ \frac{1}{2}\bigl[K^{14}+K^{23}\pm iK^{24}\mp
   iK^{13}\bigr]\\[2mm]
    K_2^\pm&\ =\ \frac{1}{2}\bigl[-K^{14}+K^{23}\mp iK^{24}\mp
iK^{13}\bigr]\ \ .
  \end{split}
\end{equation}
The elements $K_\a^\pm$ either commute with $K_1^0, K_2^0$ or shift the
corresponding eigenvalue by $\pm1$. Finally there are four fermionic
raising and four fermionic lowering operators ($\alpha$=1,2)
\begin{align}
  \label{eq:RootTwo}
  S_{1\alpha}^\pm&\ =\ S_\alpha^1\pm iS_\alpha^2&
  S_{2\alpha}^\pm&\ =\ S_\alpha^3\pm iS_\alpha^4\ \ ,
\end{align}
which raise/lower the eigenvalues of $K_1^0, K_2^0$ by $\pm 1/2$.
A complete set of relations between the new generators of the Lie
algebra $\psl$ can be read off from \eqref{pslfirst}--\eqref{psllast}
below.
\smallskip

\subsection{Kac modules and their characters}

In the present case the bosonic subalgebra $\g^{(0)}$ consists of
two commuting copies of $\sll$. The Kac modules \cite{Kac:1977em}
of $\psl$ are then labelled by pairs $(\mu,\nu)$ of representations
$\mu,\nu$ of $sl(2)$. By construction, we declare that the corresponding
representation space $V_{(\mu,\nu)}$ is annihilated by $S_1^a$ and
then generate the Kac module $[\mu,\nu]$ through application of
the raising operators $S^a_2$,
$$  [\mu,\nu] \ := \
    {\rm Ind}_{\g^{(0)} \oplus\, \g^{(1)}_+}^{\g} V_{(\mu,\nu)} \ = \
    \mc{U}(\g) \otimes_{\g^{(0)} \oplus\, \g^{(1)}_+} V_{(\mu,\nu)} \ \ .
$$
Here, we have extended the $\g^{(0)}$-module $V_{(\mu,\nu)}$ to a
representation of $\g^{(0)} \oplus \, \g_+^{(1)}$ by setting $S^a_1
V_{(\mu,\nu)} = 0$. Note that we can apply at most four fermionic
generators to the states in $V_{(\mu,\nu)}$. When choosing the
labels $\mu,\nu$ we silently agreed to identify the Cartan
subalgebra of $\psl$ with that of its maximal bosonic subalgebra.
\smallskip

To each of these Kac modules we can associate a supercharacter
according to the standard prescription\footnote{We wish to
emphasize that characters and supercharacters are related
by the substitution $z^{1/2}\to-z^{1/2}$. Hence, they both
encode precisely the same information. Some of the formulas
below possess a more natural interpretation in terms of
supercharacters though.}
\begin{equation}
  \label{eq:FullChar}
  \chi_{[\mu,\nu]}(z_1,z_2)
  \ =\ \str\Bigl[z_1^{K_1^0}\,z_2^{K_2^0}\Bigr]
  \ =\ \tr\Bigl[(-1)^Fz_1^{K_1^0}\,z_2^{K_2^0}\Bigr]
  \ =\ \chi_{\mu}(z_1)\,\chi_{\nu}(z_2)\,\chi_F(z_1,z_2)\ \ .
\end{equation}
It encodes the complete information about the weight content but
not how the vectors are linked internally. The symbols $\chi_\mu$
and $\chi_\nu$ denote $\sll$-characters while the last factor
$\chi_F$ stems from the contribution of the fermionic generators
and is independent of the choice of $\mu$ and $\nu$. For the
definition of the supercharacters we will always assume that
the ground states, i.e.\ the states in the representation
$V_{(\mu,\nu)}$ we started with, are bosonic.

In order to determine the fermionic term $\chi_F$ in the
characters \eqref{eq:FullChar} we recall that the fermionic
raising operators transform in the representation $(1/2,1/2)$ of
$\sll\oplus\sll$ while products of more than one generator
transform in antisymmetrized products thereof. This implies that
the fermionic contribution to Kac modules has the bosonic
content\footnote{Here and in the following the phrase ``bosonic
content'' refers to the decomposition of a Lie superalgebra module
with respect to the maximal bosonic subalgebra.}
\begin{equation}
  \label{eq:FermCont}
  V_F\ =\ 2(0,0)\oplus2(1/2,1/2)\oplus(1,0)\oplus(0,1)\ \ .
\end{equation}
From this we read off immediately that
\begin{equation}
  \chi_F(z_1,z_2)
  \ =\ 4+z_1+z_1^{-1}+z_2+z_2^{-1} -2(z_1^{\frac{1}{2}} +
       z_1^{-\frac{1}{2}})(z_2^{\frac{1}{2}} +
     z_2^{-\frac{1}{2}})\ \ .
\end{equation}
We introduced a special symbol for this representation since it
will appear frequently throughout the text. Note that $\chi_F =
\chi_{[0,0]}$ coincides with the character of the Kac module
generated from the trivial representation.
\smallskip

Obviously, the bosonic contributions to the characters will
strongly depend on the labels $\mu$ and $\nu$. In view of our
applications to the Lie supergroup $\PSL$ we are in fact primarily
interested in representations for which $\nu = j_2 =
0,1/2,1,\dots,$ labels the finite dimensional irreducible
representations of $\sll$ so that
\begin{equation} \label{charfd}
  \chi_\nu (z_2) \ = \ \chi_{j_2}(z_2)\ =\ \sum_{l=-j_2}^{j_2} z_2^l\ \ .
\end{equation}
The first label $\mu$, on the other hand, will be allowed to run
through three different series of representations.
\smallskip

{\em Finite dimensional representations} of $\psl$ are obtained when
we set $\mu = j_1 = 0,1/2,1, \dots$. In this case, the
contribution to the characters \eqref{eq:FullChar} is given by
$\chi_\mu(z_1) = \chi_{j_1}(z_1)$ as defined in \eqref{charfd}. Even
though such representations are not associated to unitary
representations of $\text{su}(1,1)\oplus\su$ unless $j_1=j_2=0$, finite
dimensional representations play an important role, in particular
for the boundary WZNW model.
\smallskip

Our second series of $\psl$ representations is affiliated with the
two {\em discrete series} of $\text{su}(1,1)$. In this case, the label is $\mu
= (\pm, j_1)$ with $j_1 < 0$ any negative real number.\footnote{
Our notation seems to deviate from the standard one but it
appears to be closer to the actual construction of the modules and
hence the formulas we encounter will be easier.} With our
choice of $j_1$ and of the indefinite metric \eqref{eq:Metric},
the value of the Casimir element in $(\pm,j_1)$
is given by $-j_1(j_1+1)$. By definition, the representations
$(+,j_1)$ and $(-,j_1)$ have a lowest/highest weight with $K^0_1$
eigenvalues $-j_1>0$ and $j_1<0$, respectively. The
corresponding characters are given by
\begin{equation}
    \chi_{(\pm,j_1)}(z_1)
    \ =\ \sum_{n=0}^\infty z_1^{\mp j_1\pm n}
     \ =\ \frac{z_1^{\mp j_1}}{1-z^{\pm 1}_1}\ \ .
\end{equation}
In the last two lines the geometric series expression is valid for
$|z_1|<1$ and $|z_1|>1$, respectively. Let us emphasize again that,
in our conventions, the representations $(+,j_1)$ are actually
labelled by a negative real number $j_1$ although their lowest
weight has a positive weight $-j_1$.
\smallskip

The last set of representations we need comes with the {\em
principal continuous series} of $\text{su}(1,1)$. We label such
representations by tupels $\mu = (j_1,\alpha)$ where $0 \leq \a < 1$ and
$j_1 \in\mathbb{S}=-1/2 + i \mathbb{R}$. Representations in the principal
continuous series have neither highest nor lowest weight states.
Eigenvalues of the Cartan element $K^0_1$ take values on $\a +
\mathbb{Z}$. Hence the characters of the third series read
\begin{equation}
  \chi_{(j_1,\alpha)}(z_1)\ =\ \sum_{n\in\Integer}z_1^{\alpha+n}
\ \ .
\end{equation}
Note that these characters do not depend on the spin $j_1$. Yet,
the latter determines the value $-j_1(j_1+1)$ of the quadratic
Casimir.
\medskip

The importance of Kac modules stems from the fact that they are
irreducible for generic values of the labels $\mu$ and $\nu$.
Nevertheless, for special {\em atypical} choices of $(\mu,\nu)$,
non-trivial invariant subspaces exist. A close inspection of
the action of fermionic generators on Kac modules reveals that,
starting with a bosonic {\em highest or lowest weight representation},
there is just one single atypicality condition which may be written in the
form\footnote{This conditions arises if one tries to return from
the state $S_2^1S_2^2S_2^3S_2^4|j_1,j_2\rangle$ on the fourth fermion
level to the original highest weight state $|j_1,j_2\rangle$.}
\begin{equation}
  \label{eq:AtypCond}
  0\ =\ (j_2-j_1)(j_1+j_2+1)\ =\ -j_1(j_1+1)+j_2(j_2+1) \ = \
  C_2(j_1,j_2) \ \ ,
\end{equation}
i.e.\ the Kac module $[\mu,\nu]$ possesses a non-trivial invariant
submodule if and only if the quadratic Casimir of the bosonic
subalgebra vanishes on the multiplet $(\mu,\nu)$ from which the
Kac module is generated. For finite dimensional representations
this happens whenever $j_1 = j_2$. Similarly, the Kac modules
$\bigl[(\pm,j_1),j_2\bigr]$ cease to be irreducible if and only if
$j_1=-j_2-1$. In the following we shall study the
atypical cases in much more detail.
\smallskip

In case of the principal continuous series, finally,
the atypically condition \eqref{eq:AtypCond} does not apply.
But since the value of the quadratic Casimir is directly determined
by the label of the bosonic highest weight multiplet and identical on the
whole representation generated from it we can give a
necessary condition for the decoupling of a bosonic subrepresentation
$\bigl((j'_1,\a'),j'_2\bigr)$ of the Kac module $\bigl[(j_1,\a),j_2\bigr]$:
The eigenvalues
of the Casimir operator have to agree. A careful analysis of
this condition including the discussion of possible decomposition
series then shows that the Kac modules $\bigl[(j_1,\a),j_2\bigr]$
are always irreducible.

\subsection{Finite dimensional atypical representations}

As described in the previous subsection, finite dimensional Kac
modules of $\psl$ are labelled by pairs $[j_1,j_2]$ with $j_i =
0,1/2,1,\dots$. A Kac module $[j_1,j_2]$ is irreducible whenever
$j_1 \neq j_2$. In case $j_1=j_2$, however, Kac modules turn out
to be indecomposable composites of smaller irreducible building
blocks (short multiplets). We shall describe the latter in the
next paragraph before discussing the new class of so-called
projective covers. These are maximal indecomposable composites of
short multiplets. In some sense that we shall make more precise
below, the projective covers should be considered as the natural
replacement of Kac modules in case we are dealing with atypical
representations.
\smallskip

\subsubsection{Atypical Kac modules and irreducible representations}

As we have mentioned several times, the Kac modules $[j,j]$
contain non-trivial invariant subspaces. For $j \geq 1$ the
structure of the Kac module can be encoded in the following
chain
\begin{equation}  \label{Kj}
 [j,j]:\quad [j] \ \rightarrow \  [j+\sfrac12]
   \oplus\ [j-\sfrac12]\ \rightarrow [j] \ \ ,
\end{equation}
where $[j]$ denote irreducible atypical representations
(short multiplets) of $\psl$. The structure of the reducible
Kac modules can also be depicted by a planar diagram in
which the vertical direction refers to the spin $j$ of the atypical
constituents,
\begin{equation}
  \xymatrixrowsep{8pt}\xymatrixcolsep{8pt}
 [j,j]:\quad  \xymatrix{ & [j+\foh]\ar[dr] &\\
     [j]\ar[dr] \ar[ur] && [j]\ \ . \\
             & [j-\foh] \ar[ur]&}
\end{equation}
Since pictures of this type will appear frequently throughout
this text, let us pause here for a moment and explain carefully
how to decode their information. We read the diagram \eqref{Kj}
from right to left. The rightmost entry in our chain contains
the so-called {\em socle} of the indecomposable representation,
i.e.\ the largest fully reducible invariant submodule we can find.
In the case of our Kac module, the socle happens to be irreducible
and it is given by the atypical representation $[j]$. If we
divide the Kac module by the submodule $[j]$, we obtain a new
indecomposable representation of our Lie superalgebra.
Its diagram is obtained from the one above by removing the
last entry and all arrows connected to it. The socle of this
quotient is a direct sum of the two atypical representations
$[j\pm 1/2]$.
It is rather obvious how to iterate this procedure until the
entire indecomposable representation is split up into floors
with only direct sums of irreducible representations appearing
on each floor.
\smallskip

There are two special cases for which the decomposition of the
Kac module does not follow the generic pattern as described in
eq.~\eqref{Kj}. These are the cases $j = 0$ and $j=1/2$,
\begin{align}
 [0,0]:\quad&\ [0] \ \rightarrow \  [\sfrac12]
    \rightarrow\ [0] \ \ ,\label{K0} \\[2mm]
 [\sfrac12,\sfrac12]:\quad&\ [\sfrac12] \ \rightarrow \
 [1] \ \rightarrow \ [0]\oplus [0] \ \rightarrow\ %
 [\sfrac12] \ \ .  \label{K12}
 \end{align}
The irreducible constituents $[0]$ and $[1/2]$ are the
trivial one-dimensional representation and the 14-dimensional
adjoint representation of $\psl$.
\smallskip

The supercharacters of short multiplets can be deduced from those of
the corresponding Kac modules and the composition patterns
\eqref{Kj}, \eqref{K0} and \eqref{K12}. They possess the form
\begin{equation}\label{evendec1}
\chi_{[j]}(z_1,z_2) \ = \ 2\chi_{j}(z_1) \chi_{j}(z_2)
-\chi_{j+\sfrac12}(z_1) \chi_{j-\sfrac12}(z_2)  -
\chi_{j-\sfrac12}(z_1) \chi_{j+\sfrac12}(z_2)
\end{equation}
for all $j>0$. We would like to stress that these supercharacters do
not contain the fermionic factor $\chi_F$ that appears in all
supercharacters of typical irreducible representations.

\subsubsection{Projective covers of $[j]$}

In the previous subsection we have seen the first examples of
representations which are built out of several short multiplets.
Kac modules are only one example of such composites and we shall
indeed need another class of indecomposables as we proceed, the
so-called projective covers $\P_j$. By definition, these
are the largest indecomposables whose socle consists of a single
atypical representation $[j]$. General results imply that such a
maximal indecomposable extension of $[j]$ exists and is unique
\cite{MR1378540}. In
case of $j \geq 3/2$, the structure of  $\P_j$ is encoded in
the following diagram
\begin{align} \label{Pj}
  \P_j:  &\ [j] \longrightarrow
             2 [j+\sfrac12] \oplus 2 [j-\sfrac12]
                \longrightarrow
              [j+1]\oplus 4[j] \oplus
               [j-1] \longrightarrow \\[2mm]
           & \hspace*{4cm} \longrightarrow
                2 [j+\sfrac12] \oplus 2 [j-\sfrac12]
               \longrightarrow [j] \ \ . \nonumber
\end{align}
Note that $\P_j$ contains an entire Kac module as a proper
submodule. In this sense, the Kac modules are extendable. We
also observe one rather generic feature of projective covers:
they are built up from different Kac modules in a way that
resembles the pattern in which Kac modules are constructed
out of irreducibles (see eq.\ \eqref{Kj}).\footnote{In
mathematics this statement is known as a generalization of
the BGG duality theorem \cite{MR1378540}.} One may see this
even more clearly if $\P_j$ is displayed as a 2-dimensional
diagram in which the additional direction keeps track of
the spin $j$ of the atypical constituents $[j]$,
\begin{equation}
  \xymatrixrowsep{8pt}\xymatrixcolsep{8pt}
  \P_j: \
  \xymatrix{& &  [j+1] \ar[dr]& &\\
 & 2[j+1/2]\ar[dr] \ar[ur] & & 2[j+1/2]\ar[dr]&\\
     [j]\ar[dr] \ar[ur] && 4[j]\ar[ur]\ar[dr] & &[j]\ . \\
             & 2[j-1/2] \ar[ur]\ar[dr]&&2[j-1/2]\ar[ur] &\\
& & [j-1] \ar[ur] &}
\end{equation}
We will continue to switch between such planar pictures and
diagrams of the form \eqref{Pj}. The remaining cases $j=0,1/2,1$
have to be listed separately. When $j=1$ the picture is very
similar only that we have to insert $2[0]$ in place of
$[j-1]$,
\begin{equation} \label{P1}
  \P_1:  \   [1] \longrightarrow
               2 [ \fth ]
                  \oplus 2 [\sfrac12]
               \longrightarrow  [2]\oplus 4[1] \oplus
               2 [0] \longrightarrow
                2 [ \fth ]
                  \oplus 2 [\sfrac12]
               \longrightarrow [1] \ \ .
\end{equation}
The projective cover of the atypical representation $[1/2]$ is
obtained from the generic case by the formal substitution $2[j-1/2]
\rightarrow 3[0]$,
\begin{equation} \label{P12}
  \P_{\frac{1}{2}}: \   [\sfrac12] \longrightarrow
           2 [1] \oplus 3 [0] \longrightarrow
              [ \fth ]
                  \oplus 4 [\sfrac12]
               \longrightarrow   2 [1] \oplus 3 [0]
               \longrightarrow [\sfrac12] \ \ .
\end{equation}
Finally, the projective cover $\P_0$ of the trivial
representation is given by,
\begin{equation} \label{P0}
  \P_0:  \  [0] \longrightarrow 3 [\sfrac12]
           \longrightarrow 2 [1] \oplus 6 [0]
               \longrightarrow 3 [\sfrac12]
             \longrightarrow [0] \ \ .
\end{equation}
The reader is invited to convert the last three formulas into
planar pictures.
\smallskip

This concludes our list of the projective covers of finite dimensional
representations. The representations $\P_j$ appear in the operator
products of certain open string vertex operators in the WZNW model,
when we consider boundary conditions corresponding to a point-like
brane. Together, typical representations and the projective covers
of atypicals form the subset of so-called projective
representations. What makes this class particularly interesting is
its behavior under tensor products. In fact, it is well-known that
projective representations of a Lie superalgebra form an ideal in
the fusion ring. This means that the product of a projective
representation with any other representation, no matter how
complicated it is, can be decomposed into projectives. We shall
later see that this property of projective representations (along
with the fact that they are easy to list) has invaluable
consequences.
\smallskip

It is moreover relevant to observe that, unlike for
the atypicals themselves, the characters of their projective covers
contain the full fermionic character $\chi_F$ as a factor. To be
precise one has
\begin{equation}
  \label{eq:ProjBos}
  \chi_{\mc{P}_j}
  \ =\ \Bigl[2\chi_j(z_1)\chi_j(z_2)-\chi_{j+\frac12}(z_1)
   \chi_{j+\frac12}(z_2) -\chi_{|j-\frac12|}(z_1)
\chi_{|j-\frac12|}(z_2)\Bigr]\,\chi_F(z_1,z_2)\ .
\end{equation}
This property puts projective covers on an equal footing with
typical irreducibles. Eventually, it will even allow us to come up
with a version of the familiar Racah-Speiser algorithm which
holds for projective representations of Lie superalgebras.
We refer readers interested in further details to section
\ref{sc:Branching} below.

\subsection{Infinite dimensional atypical representations}

Let us now turn to the theory of infinite dimensional atypical
representations of $\psl$. As we have remarked before, atypicals
appear only in the discrete series and if the labels $(\mu,\nu)=
\bigl((\pm,j_1),j_2\bigr)$ satisfy the condition $j_1+j_2+1=0$. The plan of
this subsection follows the same logic as our discussion of finite
dimensional atypicals, i.e.\ we shall study atypical Kac modules
and irreducibles first and then turn to the projective covers. But
since some of the results below seem to be less known, we will be
a bit more detailed about their derivation.

\subsubsection{Atypical Kac modules and irreducible representations}

Kac modules of the form $\bigl[(\pm,-j-1),j\bigr]$ fail to be irreducible.
In order to understand the structure of the resulting Kac modules
let us first have a look at the bosonic content of typical
representations,
\begin{equation}
  \bigl[(\pm,-j-1),j\bigr]\bigr|_{\sll\oplus\sll}
  \ =\ \bigl((\pm,-j-1),j\bigr)\otimes V_F\ \ ,
\end{equation}
where $V_F$ denotes the fermionic contributions as specified in
\eqref{eq:FermCont}. This tensor product can be evaluated using
the familiar rules for $sl(2)$ and the additional formula
\begin{equation}
  \label{eq:TP}
  (\pm,-j-1)\otimes k\ =\ \bigoplus_{l=-j-1-k}^{-j-1+k}\,(\pm,l)\ \ ,
\end{equation}
which holds as long as $j\geq0$ and $k \leq j$, or more generally as
long as the sum on the right hand side does not contain
contributions with non-negative half-integer or integer $l$. It is
straightforward to see that in $\bigl[(\pm,-j-1),j\bigr]$ the bosonic
representations $\bigl((\pm,-j-1),j\bigr)$, $\bigl((\pm,-j-\foh),j-\foh\bigr)$ and
$\bigl((\pm,-j-\fth),j+\foh\bigr)$ decouple. After dividing out the
induced invariant submodules we are left with the irreducible
representation $[j]_\pm$ whose bosonic content reads
\begin{equation} \label{j-dec}
  [j]_\pm\bigr|_{\sll\oplus\sll}\ = \
  2\bigl((\pm,-j-1),j\bigr)
  \oplus\bigl((\pm,-j-\fth),j-\foh\bigr)
  \oplus\bigl((\pm,-j-\foh),j+\foh\bigr)
\end{equation}
for $j\neq 0$. We also infer that the structure of the degenerate
Kac modules is described by the composition series
\begin{equation} \label{-jdec}
  \bigl[(\pm,-j-1),j\bigr]:\qquad[j]_\pm\to[j+\foh]_\pm\oplus[j-\foh]_\pm\to[j]_\pm\ \ .
\end{equation}
Again, we assumed that $j\neq 0$. Formally this formula is
identical to the one which is obtained for finite dimensional
representations \cite{Gotz:2005ka}.
\smallskip

So far we have avoided to investigate the special case $j=0$. The
characters of the irreducible representations $[0]_\pm$ are easily
obtained from the above since these representations arise as
building blocks of the Kac modules $\bigl[(\pm, -3/2),1/2\bigr]$,
\begin{equation}
  [0]_\pm\bigr|_{\sll\oplus\sll} \ = \
 2 \bigl((\pm,-1),0\bigr)\oplus\bigl((\pm,-1/2),1/2\bigr) \ \ .
\end{equation}
Concerning the structure of the special Kac modules $\bigl[(\pm,-1),0\bigr]$
we note that their bosonic content is given by
\begin{equation}
  \begin{split}
    \bigl[(\pm,-1),0\bigr]\bigr|_{\sll\oplus\sll}
    &\ =\ 2\bigl((\pm,-1),0\bigr)\oplus2\bigl((\pm,-3/2),1/2\bigr)
     \oplus2\bigl((\pm,-1/2),1/2\bigr)\\[2mm]
    &\qquad\oplus\mc{R}_{((\pm,-1),0)}\oplus\bigl((\pm,-2),0\bigr)
    \oplus\bigl((\pm,-1),1\bigr)\ \ .
  \end{split}
\end{equation}
Let us stress that it contains an indecomposable bosonic
representation $\mc{R}_{((\pm,-1),0)}$ which has the decomposition
series
\begin{equation}
  \mc{R}_{((\pm,-1),0)}:\qquad\bigl((\pm,-1),0\bigr)\to(0,0)\to
 \bigl((\pm,-1),0\bigr)\ \ .
\end{equation}
The structure of the Kac module may be summarized in the
decomposition series
\begin{equation} \label{-0dec}
  \bigl[(\pm,-1),0\bigr]:\qquad[0]_\pm\to[0]\oplus[1/2]_\pm\to[0]_\pm\ \ .
\end{equation}
It is interesting to find a finite dimensional representation in
the decomposition series even though we started with an infinite
dimensional representation.
\smallskip

For later applications we shall draw an important conclusion from
the decomposition formulas \eqref{-jdec} and \eqref{-0dec} of Kac
modules. Note that they allow us to express the supercharacter of
the atypical trivial representation $[0]$ formally as in infinite
sum over supercharacters of Kac modules,\footnote{We have first
learned this trick and its generalization to affine
supercharacters from Hubert Saleur \cite{Saleur:2005}, see also
\cite{Rozansky:1992td} for a very simple version thereof.}
\begin{equation} \label{0series}
\chi_{[0]}(z_1,z_2) \ = \  \sum_{n=0}^\infty \ (n+1) \,
\chi_{[(\pm,-n/2-1),n/2]}(z_1,z_2) \ \ .
\end{equation}
Indeed, one can show by straightforward direct computation that
the terms on the right hand side sum up to $\chi_{[0]} =1$. If
the supercharacters of the Kac modules are decomposed into a sum
of bosonic characters as encoded in  formulas \eqref{-jdec} and
\eqref{-0dec}, then all but the contribution from the trivial
representation cancel each other.

\subsubsection{Projective covers of $[j]_\pm$}

In case of finite dimensional representations, the projective
covers are built up from Kac modules and there exists a rather
simple rule to determine the number of Kac modules of any type
within a given projective cover \cite{MR1378540}. If we
extrapolate this rule to our present setup, we arrive at the
following composition series for the projective covers of the
discrete representations
\begin{align}
\P_j^\pm:&\ [j]_\pm \longrightarrow 2 [j+\foh]_\pm \oplus
2 [j-\foh]_\pm \longrightarrow [j+1]_\pm \oplus 4 [j]_\pm \oplus
[j-1]_\pm \nonumber \\[2mm] & \hspace*{2cm}
\longrightarrow 2 [j+\foh]_\pm \oplus 2[j-\foh]_\pm \longrightarrow
[j]_\pm \nonumber
\end{align}
for $j\geq1$. The same expression may be used for $j=1/2$ if we formally
replace $[-\foh]_{\pm}$ by the trivial representation $[0]$.
The structure of the projective covers of
$[0]_\pm$ does not follow the generic pattern. Instead it is given
by
$$
\P_0^\pm  \ \ : \ \ [0]_\pm \longrightarrow [0] \oplus 2
[\foh]_\pm \longrightarrow 3 [0]_\pm \oplus [1]_\pm \longrightarrow
[0] \oplus 2 [\foh]_\pm \longrightarrow [0]_\pm \ \ .
$$
Needless to stress that the characters of these projective covers
contain the factor $\chi_F$ as in the finite dimensional case.

\newpage
\section{\label{sc:Representations}Representation theory of the
affine superalgebra}

Irreducible representations of the affine $\psl$ superalgebra can be
built over all the irreducible representations of the $\psl$
algebra that we discussed above. The latter can be either finite
or infinite dimensional. We shall address the infinite dimensional
ones first and then turn to the finite dimensional representations
in the second subsection.

\subsection{The affine $\psl$ algebra and spectral flows}

Here we display the definition of the current algebra $\apsl(2|2)_k$
first. In terms of raising and lowering operators
\eqref{eq:RootOne}--\eqref{eq:RootTwo} it may be written as
\begin{align}\label{pslfirst}
   [K_{1,m}^0,K_{1,n}^\pm]&\ =\ \pm\,K_{1,m+n}^\pm &
   [K_{2,m}^0,K_{2,n}^\pm]&\ =\ \pm\,K_{2,m+n}^\pm\\[2mm]
   [K_{1,m}^0,S_{1\alpha,n}^\pm]&\ =\ \pm\frac{1}{2}\,S_{1\alpha,m+n}^\pm&
   [K_{1,m}^0,S_{2\alpha,n}^\pm]&\ =\
\pm\frac{1}{2}\,S_{2\alpha,m+n}^\pm\\[2mm]
   [K_{2,m}^0,S_{1\alpha,n}^\pm]&\ =\ \pm\frac{1}{2}\,S_{1\alpha,m+n}^\pm&
   [K_{2,m}^0,S_{2\alpha,n}^\pm]&\ =\
\mp\frac{1}{2}\,S_{2\alpha,m+n}^\pm\\[2mm]
   \{S_{1\alpha,m}^\pm,S_{2\beta,n}^\pm\}&\ =\ \mp
2\epsilon_{\alpha\beta}\,K_{1,m+n}^\pm&
   \{S_{1\alpha,m}^\pm,S_{2\beta,n}^\mp\}&\ =\ \pm
2\epsilon_{\alpha\beta}\,K_{2,m+n}^\pm\\[2mm]
   [K_{1,m}^\pm,S_{1\alpha,n}^\mp]&\ =\ \pm S_{2\alpha,m+n}^\pm&
   [K_{1,m}^\pm,S_{2\alpha,n}^\mp]&\ =\ \mp S_{1\alpha,m+n}^\pm\\[2mm]
   [K_{2,m}^\pm,S_{1\alpha,n}^\mp]&\ =\ \pm S_{2\alpha,m+n}^\mp&
   [K_{2,m}^\pm,S_{2\alpha,n}^\pm]&\ =\ \mp S_{1\alpha,m+n}^\pm\ \ .
\end{align}
In addition, there are six relations involving the level $k$ of the
$\psl$ current algebra. These read as follows,
\begin{align}
  [K_{1,m}^0,K_{1,n}^0]&\ =\ -\frac{k}{2}\,m\,\delta_{m+n,0}\hspace{2.5cm}
  [K_{2,m}^0,K_{2,n}^0]\ =\ \frac{k}{2}\,m\,\delta_{m+n,0}\\[2mm]
  [K_{1,m}^+,K_{1,n}^-]&\ =\ 2K_{1,m+n}^0 -mk\,\delta_{m+n,0}\hspace{1cm}
   [K_{2,m}^+,K_{2,n}^-]\ =\ 2K_{2,m+n}^0 +mk\,\delta_{m+n,0}\\[2mm]
   \{S_{1\alpha,m}^+,S_{1\beta,n}^-\}&\ =\ %
2\epsilon_{\alpha\beta}\bigl(K_{1,m+n}
^0-K_{2,m+n}^0\bigr)- 2\,mk\,\epsilon_{\alpha\beta}\,\delta_{m+n,0}\\[2mm]
   \{S_{2\alpha,m}^+,S_{2\beta,n}^-\}&\ =\ %
2\epsilon_{\alpha\beta}\bigl(K_{1,m+n} ^0+K_{2,m+n}^0\bigr) -
2\,mk\,\epsilon_{\alpha\beta}\,\delta_{m+n,0}\label{psllast}\ \ .
\end{align}
The algebra defined by eqs. \eqref{pslfirst}--\eqref{psllast}
possesses a two-parameter family $\c^{(w_1,w_2)}$ of
automorphisms. It is induced from the following two-parameter
family of automorphisms for the bosonic subalgebra
$\asl_{-k}\oplus\asl_k$
\begin{align}
   \c^{(w_1,w_2)} (K_{1,n}^0)&\ =\  K_{1,n}^0-\frac{k}{2}\,w_1\,
   \delta_{n0}&
   \c^{(w_1,w_2)} (K_{1,n}^\pm)&\ =\  K_{1,n\pm w_1}^\pm\label{antiK}
    \\[2mm]
   \c^{(w_1,w_2)} (K_{2,n}^0)&\ =\  K_{2,n}^0+\frac{k}{2}\, w_2 \,
    \delta_{n0}& \c^{(w_1,w_2)} (K_{2,n}^\pm)&\ =\  K_{2,n\pm w_2}^\pm
\label{antiK2}\ \ .
\end{align}
One may easily check that these maps extend to the whole algebra
$\apsl(2|2)_k$ through
\begin{align}
\c^{(w_1,w_2)}(S_{1\alpha,n}^\pm) &\ =\
 S_{1\alpha,n\pm\frac{w_1+w_2}{2}}^\pm \label{antiS}&
 \c^{(w_1,w_2)} ( S_{2\alpha,n}^\pm) &\ =\
   S_{2\alpha,n\pm\frac{w_1-w_2}{2}}^\pm\ \ .
\end{align}
Application of these automorphisms maps representations of the
current algebra onto each other. They therefore play a crucial
role in the representations theory of $\apsl(2|2)_k$.

\subsubsection{\label{sc:FreeField}Free field construction of the affine algebra}

Following \cite{Bars:1990hx,Berkovits:1999im} we can construct
$\apsl(2|2)_k$ out of four pairs of fermions $p^a$ (of weight $h=1$)
and $\theta^a$ (with $h=0$) which satisfy
\begin{equation}\label{OPEfermion}
   p^a(z)\,\theta^b(w)\ =\ \frac{\delta^{ab}}{z-w}\ \ .
\end{equation}
and the affine $\sll_{-k-2}\oplus\sll_{k-2}$ algebra that is generated
by currents $j^{ab}$ with the following operator product
expansion
\begin{equation}\label{OPEso4}
   j^{ab}(z)\,j^{cd}(w)
   \ =\ -
\frac{k\,\epsilon^{abcd}+2(\delta^{ac}\delta^{bd}-
\delta^{ad}\delta^{bc})}{(z-w)^2}
        +\frac{i\bigl[\delta^{ac}j^{bd}-\delta^{ad}j^{bc}-
    \delta^{bc}j^{ad}+\delta^{bd}j^{ac}\bigr]}{z-w}\ \ .
\end{equation}
Based on this structure we now obtain the $\apsl(2|2)_k$
algebra via
\begin{equation}\label{K}
\begin{split}
   K^{ab} & =\ j^{ab}-i\bigl[\theta^ap^b-\theta^bp^a\bigr]\quad
   , \quad\quad S_2^a\ =\ p^a\\[2mm]
   S_1^a& =\ k\,\partial\theta^a+\frac{i}{2}\,\epsilon^{abcd}\,
    \bigl[\theta^bj^{cd}-
i\theta^b\theta^cp^d\bigr]\ \ .
  \end{split}
\end{equation}
It is straightforward to show that the fields $K$ and $S_\a^a$
obey the correct relations following from eqs.~\eqref{eq:CR}
and~\eqref{eq:Metric}.
\smallskip

  One of the main ingredients in the representation theory
  of $\apsl(2|2)_k$ are the spectral flow automorphisms which have
  been defined in eqs.~\eqref{antiK}-\eqref{antiS}. We would like
  to show that the spectral flows of the full affine Lie superalgebra
  are inherited from those which may be defined for
  $\sll_{-k-2}\oplus\sll_{k-2}$ and the fermions $p_a$ and $\theta^a$.
  For the bosonic current algebra we have the standard transformations
\begin{align}
  \label{eq:SF}
 \c^{(w_1,w_2)}(J_{1,n}^0)&=\  J_{1,n}^0-\frac{k+2}{2}\,w_1\,\delta_{n0}&
 \c^{(w_1,w_2)}(J_{1,n}^\pm)&=\  J_{1,n\pm w_1}^\pm\\[2mm]
 \c^{(w_1,w_2)}(J_{2,n}^0)&=\  J_{2,n}^0+\frac{k-2}{2}\,w_2\,\delta_{n0}&
 \c^{(w_1,w_2)}(J_{2,n}^\pm)&=\  J_{2,n\pm w_2}^\pm\ \ .
\end{align}
  In view of the schematic structure $K=J+(p\theta)$ of the currents
  appearing in the full superalgebra we also need to implement a
  non-trivial transformation on the fermions. A suitable basis is
\begin{align}
p_1^{\pm}&=p^1\pm ip^2=S_{12}^{\pm}&
p_2^{\pm}&=p^3\pm ip^4=S_{22}^{\pm}\\[2mm]
\theta_1^{\pm}&=\theta^1\pm i\theta^2&
\theta_2^{\pm}&=\theta^3\pm i\theta^4\ \ .
\end{align}
Starting from the defining relation $\left\{p^a_m,\theta^b_n
\right\}=\delta^{ab}\delta_{m+n,0}$ there remain the following
non-vanishing anti-commutators
\begin{align}\label{comptheta}
\left\{p^{\pm}_{1,m},\theta^{\mp}_{1,n} \right\}&=2\delta_{m+n,0}&
\left\{p^{\pm}_{2,m},\theta^{\mp}_{2,n} \right\}&=2\delta_{m+n,0}\ \ .
\end{align}
Employing \eqref{antiK} and \eqref{antiS}, a natural and consistent candidate
for the spectral flow on the fermions is
\begin{align}\label{sfonp}
\c^{(w_1,w_2)}(p_{1,n}^\pm)&\ =\  p_{1,n\pm\frac{w_1+w_2}{2}}^\pm &
\c^{(w_1,w_2)}(p_{2,n}^\pm)&\ =\  p_{2,n\pm\frac{w_1-w_2}{2}}^\pm\\[2mm]
\c^{(w_1,w_2)}(\theta_{1,n}^\pm)&\ =\  \theta_{1,n\pm\frac{w_1+w_2}{2}}^\pm &
\c^{(w_1,w_2)}(\theta_{2,n}^\pm)&\ =\  \theta_{2,n\pm\frac{w_1-w_2}{2}}^\pm\ \ .
\end{align}
One can easily check that this transformation leaves \eqref{comptheta}
invariant. The consistency of the action of $\c^{(w_1,w_2)}$ on $j^{ab}$
and the fermions $p$ and $\theta$ with the spectral flow on $K_{1,n} ^{\pm}$,
 $K_{1,n} ^{\pm}$, $K^0_{1,m}$ and  $K^0_{2,m}$ given in~\eqref{antiK}
and \eqref{antiK2} can be verified by expressing the latter in
terms of the $(p^{\pm},\theta^{\pm})$-basis and taking care about
normal ordering.

\subsection{\label{sc:AffineReps}Typical representations and their supercharacters}

In the following we shall be concerned with the various
representations of the affine algebra $\apsl(2|2)_k$. Without
further mentioning we shall always assume that the level $k$ is an
integer $k \geq 3$. The standard representations are obtained by
acting with generators of negative mode number $n$ on ground
states which can transform in the various representations of the
zero mode algebra $\psl$. Hence, generic representations
$[\mu,\nu]^\wedge$ of the current algebra are labelled by the same
pairs of sl(2) representation labels $\mu,\nu$ as the modules
$[\mu,\nu]$ of $\psl$, with some additional $k$-dependent
truncations on the possible range of the spin labels.

As before we take the second label $\nu = j_2 = 0, 1/2,1, \dots$
to be a half-integer. We shall also require that $j_2 \leq k/2-1$.
As for the first label, there are again three different series
corresponding to $\mu = (\pm,j_1)$ (for $-\frac{1}{2}>j_1>-\frac{k+1}{2}$)
or $\mu = (j_1,\alpha)$ (for $j_1\in\mathbb{S}=-\frac{1}{2}+i\Real$)
for representations with an infinite number of ground states and to
$\mu = j_1 = 0,1/2,1,k/2+1$ when the number of ground states is
finite. In terms of these labels, the ground states of the
associated representations possess conformal dimension\footnote{
We will omit the hat in affine representations if this interpretation
is clear from the context.}
$$ h_{[(\pm, j_1),j_2]} \ = \ \bigl(- j_1(j_1+1)+j_2(j_2+1) \bigr)/k\ \  . $$
Our aim is to describe the singular vectors in the Verma modules
and to provide the associated formulas for the supercharacters
$$ \chi_{R}(q,z_1,z_2) \ := \ \str_{R}
   \left(\, q^{L_0-\frac{c}{24}}\, z_1^{K^0_1}\, z_2^{K^0_2}\,
\right)
$$
of irreducible representations since these are the basic building
blocks of any type of representation.
We define typical representations to be those which result from
Verma modules in which all the singular vectors are
inherited from the bosonic subalgebra. As a consequence, these
typical representations have a very nice representation in terms
of the free field construction \eqref{K}. In order to clarify this statement
we recall that the pairs of fermions have a unique representation
$\mc{F}$ if we restrict ourselves to integer moding.\footnote{With
non-integer moding global supersymmetry cannot be realized in the
WZNW model we are aiming at.} Given any irreducible representation
$\mc{V}_{(\mu,\nu)}$ of $\asl_{-k-2}\oplus\asl_{k-2}$
we may then define an action of $\apsl(2|2)_k$ on the
generalized Fock module
\begin{equation}
  \label{eq:AffKac}
  \mc{V}_{[\mu,\nu]}\ =\ \mc{V}_{(\mu,\nu)}\otimes\mc{F}
\end{equation}
  using the free fermion realization. Our terminology ensures that
  this representation is irreducible if and only if it is typical.
  This follows from the analysis of the Kac-Kazhdan determinant,
  see appendix \ref{sc:KacKazhdan}. Hence, the supercharacter of an
  irreducible typical representation takes the form
\begin{align}  \label{chargentyp}
\chi_{[\mu,\nu]}(q,z_1,z_2) & =  \frac{1}{\eta^{4}(q)}
  \, \prod_{a,b = \pm1} \, \vartheta_1(z_1^{a/2}
 z_2^{b/2},q) \ \chi^{-k-2}_{\mu}(q,z_1)\ \chi^{k-2}_{\nu}(q,z_2)
 \\[2mm]  &  \hspace*{-3cm}  \text{ where} \ \
\vartheta_1(y,q) \ = \ -i y^{1/2} q^{1/8}\  \prod_{n=1}^\infty \,
(1-q^n) (1-yq^n) (1-y^{-1} q^{n-1})\ \ .
\end{align}
The first factors $\vartheta_1/\eta$ in the character
\eqref{chargentyp} arise from the four pairs of fermionic fields
$p_a$ and $\theta^a$ in the free field construction.
\smallskip

Atypical Kac modules are obtained when $j_1-j_2=nk$ or $j_1+
j_2+1=nk$ for some $n\in\Integer$ and they possess additional
singular vectors resulting from the application of fermionic generators,
see appendix \ref{sc:KacKazhdan}. Given the physical $k$-dependent
bounds on the spins it is easy to see that the only atypicality
conditions which apply are $j_1=j_2$ and $j_1+j_2+1=0$, just as in the
zero mode sector. Hence all the affine submodules originate
from singular vectors on the level of the ground states.
The corresponding irreducible representations of
$\apsl(2|2)_k$ are denoted by $[j]^\wedge$ and $[j]_\pm^\wedge$,
respectively. In the following we shall present explicit formulas
for the supercharacters of all these representations.

\subsection{Representations with an infinite number of ground states}

To begin with we shall present formulas for the supercharacters of
the discrete and principal continuous series representations of
the affine algebra $\apsl(2|2)_k$. In the former series special
attention will be paid to the atypical cases. In this sector
the characters may be expressed in terms of infinite sums of
characters of typical representations, see \cite{Rozansky:1992td}
for the first formula of this type in the context of $\gl$ and
unpublished work of Hubert Saleur \cite{Saleur:2005} for a more
elaborate application.

\subsubsection{Typical discrete series representations}

Physically relevant typical representations of the discrete series
are labelled by $j_1$ and $j_2$ with $j_1 + j_2+1 \neq0$ and $j_1 <
-1/2$. We shall also keep our restriction to values $j_1 \geq
-(k+1)/2$. Furthermore, $j_2$ are certainly taken from the usual set
$j_2 = 0,1/2,1, \dots, k/2-1$. The characters of the corresponding
representations of $\apsl(2|2)_k$ read,
\begin{equation}  \label{charinfintyp}
\chi_{[(\pm,j_1),j_2]}(q,z_1,z_2) \ = \ \frac{1}{\eta^{4}(q)}
  \, \prod_{\nu,\mu = \pm} \, \vartheta_1(z_1^{\nu /2}
 z_2^{\mu /2},q) \ \chi^{-k-2}_{(\pm,j_1)}(q,z_1)\ \chi^{k-2}_{j_2}(q,z_2)
\end{equation}
where $j_1 + j_2+1 \neq 0$. Let us recall
that the relevant $\asl$ characters are given by
\begin{align}
   \chi^{-k-2}_{(\pm,j_1)} (q,z_1) &\ =\ \pm i q^{-\frac{(j_1+1/2)^2}{k}}
   \, z_1^{\mp(j_1+1/2)} \,
   \vartheta_1(q,z_1)^{-1}  \label{sl2char}\\[2mm]
   \chi^{k-2}_{j_2} (q,z_2) &\ =\ %
   i q^{\frac{(j_2+1/2)^2}{k}} z_2^{j_2+1/2} \,
   \vartheta_1(q,z_2)^{-1} \ \Psi^k_{j_2}(q,z_2) \label{su2char}\\[4mm]
  \text{where} \ \  \Psi^k_{j_2}(q,z_2) &\ =\ %
 \sum_{a\in\Integer}\, q^{ka^2+2a(j_2+1/2)} \left(z_2^{ak}-z_2^{-ak-2(j_2+1/2)}
   \right)\ \ .
\end{align}
We amend eq.\ \eqref{sl2char} by the prescription to
expand the function $\vartheta$ in powers of $z_1$ for
the positive sign and in powers of $1/z_1$ for the
negative sign. Similarly, an expansion in $1/z_2$ is
to be used in equation \eqref{su2char}. In the
following we shall use the symbol $\bigl[(\pm,j_1),j_2\bigr]^\wedge$
for these irreducible representations of the affine algebra $\apsl(2|2)_k$.

\subsubsection{Atypical discrete series representations}

Nothing keeps us from evaluating the characters we introduced in
the previous subsection at the special points where $j_1 +j_2 +1 =
0$. At these points, the Verma modules over the representations
$\bigl[(\pm,-j-1),j\bigr]$ develop new fermionic singular vectors. The latter
are all to be found among the ground states so that the composition
of the atypical module from irreducibles is identical to the one
for the corresponding Kac modules of the horizontal subalgebra
$\psl$. For the characters this implies
\begin{equation}\label{adatyp}
\begin{split}
\chi_{[(\pm,j+1),j]}(q,z_1,z_2) & = \
   2\chi_{[j]_\pm}(q,z_1,z_2)
  - \chi_{[j+1/2]_\pm}(q,z_1,z_2)
  - \chi_{[j-1/2]_\pm}(q,z_1,z_2) \\[2mm]
\chi_{[(\pm,1),0]}(q,z_1,z_2) & =\
   2\chi_{[0]_\pm}(q,z_1,z_2)
  - \chi_{[1/2]_\pm}(q,z_1,z_2)
  - \chi_{[0]}(q,z_1,z_2)\ \ .
\end{split}
\end{equation}
Note that the character $\chi_{[0]}$ that appears in the last line
is the character of the vacuum representation. The formulas
\eqref{adatyp} can be used to solve for the characters of atypical
representations from the discrete series. In fact, formula
\eqref{0series} suggests that
\begin{align}
\chi_{[j]_\pm}(q,z_1,z_2) &\ =\ -\sum_{n=0}^\infty \
   (n+1) \, \chi_{[(\pm,-(j+3/2+n/2)),j+1/2+n/2]}(q,z_1,z_2)\nonumber \\[2mm]
\label{datyp}  &\ = \ \pm\frac{1}{\eta^4(q)\vartheta_1(q,z_1)
\vartheta_1(q,z_2)}
  \,  \prod_{\nu,\mu = \pm} \,
  \vartheta_1(z_1^{\nu /2} z_2^{\mu /2},q) \, \Psi^\pm_j(q,z_1,z_2) \\[2mm]
\text{with} \ \ \Psi_j^\pm(q,z_1,z_2) &\ =\
 \, \sum_{a\in\Integer} \, q^{ka^2 + 2a(j+1)} \,
   \left( \frac{  z_1^{\pm j \pm 1}z^{ak+j+1}_2}{(1-q^az_1^{\pm 1/2}z_2^{1/2})^2}
   -  \frac{  z_1^{\pm j \pm 1}z^{-ak- (j+1)}_2}{(1-q^az_1^{\pm 1/2}z_2^{-1/2})^2}
   \right) \ \ . \nonumber
\end{align}
One may check that these characters indeed obey the relations
\eqref{adatyp}. In view of the equation \eqref{0series} we can also formally
use the previous formula for the value $j=-1/2$ to determine the vacuum
character of the $\psl$ current algebra
\begin{align}
\chi_{[0]}(q,z_1,z_2) &\ =\ -\frac{1}{\eta^4(q)\vartheta_1(q,z_1)
\vartheta_1(q,z_2)}
  \,  \prod_{\nu,\mu = \pm} \,
  \vartheta_1(z_1^{\nu /2} z_2^{\mu /2},q) \, \Psi_0(q,z_1,z_2)
  \label{vacchar} \\[2mm]
\text{with} \ \ \Psi_0 &\ =\ %
  \sum_{a\in\Integer}\, q^{ka^2 + a} z_1^{-1/2} \,
   \left( \frac{z^{ak+1/2}_2}{(1-q^az_1^{-1/2}z_2^{1/2})^2}
     -\frac{z^{-ak-1/2}_2}{(1-q^az_1^{-1/2}z_2^{-1/2})^2} \right)
\ \ . \nonumber
\end{align}
This should be considered as an expansion in $1/z_1$ and $1/z_2$.
Note that $\Psi_0 = \Psi^+_{-1/2} = \Psi^-_{-1/2}$. The functions
$\Psi$ can be expressed as a derivative of higher level Appell
functions \cite{Kac:2001}, as in the case of $\slto$ \cite{Semikhatov:2003uc}.
For an explicit evaluation we proceed as follows: We first divide the sum over
$a\in\Integer$ in two domains, $a\geq0$ and $a<0$. For $a\geq0$
we expand the denominator as it stands while for $a<0$ we first multiply
both numerator and denominator with $q^{-2a}$ in order to
obtain an expansion in positive powers of $q$. The resulting expressions
have been verified to reproduce the structure of singular
vectors of the module $[0]^\wedge$ for $k=1,2,\dots,7$ and energies
smaller or equal to $h=4$.
\bigskip

The formulas we have proposed also pass some more general
non-trivial consistency checks. To begin with there is a
simple relation between the characters of representations
from the discrete series,
$$
\chi_{[(\pm,j_1),j_2]}(q,z_1,z_2)\ = \
    \chi_{[(\mp,j_1),j_2]}(q,z^{-1}_1,z_2)\ \ , \ \
\chi_{[j]_\pm}(q,z_1,z_2) \ = \
                     \chi_{[j]_\mp}(q,z_1^{-1},z_2) \ \ .
$$
This property expresses a manifest symmetry of the corresponding
representations. In fact, under the reflection $K^0_1 \rightarrow
- K^0_1$ and an corresponding action on the fermions which
promotes this
transformation into an automorphism, the representations from
the two different discrete series are mapped onto each other.
\smallskip

Our second crucial observation concerns the behavior of the
characters under spectral flow. We consider the spectral flow
automorphisms $\c_\pm  = \c^{(\mp 1,1)}$. Note that these generate
all spectral flows that do not interpolate between Ramond and
Neveu-Schwarz representations, i.e.\ that map integer mode numbers
to integer mode numbers. On the zero modes $L_0, K^0_1\equiv
K^0_{1,0}$ and $K^0_2\equiv K^0_{2,0}$ they act according to
$$
\c_\pm(L_0) \ = L_0 + K^0_2 \mp K^0_1 \ \ \ , \ \ \c_\pm(K^0_1) \
= \ K^0_1 \pm k/2 \ \ \ , \ \ \c_\pm (K^0_2) \ = \ K^0_2 + k/2\ \
.
$$
On characters of representations, the action of the spectral flow
can be expressed as
$$ \c_\pm \bigl( \chi_\ast(q,z_1,z_2) \bigr)
   \ = \ z_1^{\pm k/2} z_2^{k/2}\,
      \chi_\ast(q,q^{\mp 1}z_1, q z_2) \ \ .
$$
It is rather easy to see that this action is consistent with the
following behavior of irreducible representations,
$$ [(\pm,j_1),j_2]^\wedge \ \xrightarrow{\,\c_\mp\,}
   [(\mp,-k/2-1-j_1,k/2-1-j_2]^\wedge\ \ \ , \ \ \
   [j]_\pm^\wedge \ \xrightarrow{\,\c_\mp\,} \
   [k/2-1-j]_\mp^\wedge\ \ .
$$
Combining all these observations we obtain the following equation
in the case that the level $k$ is even
$$ z_1^{k/2} z_2^{k/2}\,
      \chi_{[k/4-1/2]_-}(q,q^{-1} z_1, q z_2)
  \ = \ \chi_{[k/4-1/2]_+}(q,q^{-1} z_1, q z_2) \ = \
            \chi_{[k/4-1/2]_-}(q,z_1^{-1},z_2)\ \ .
$$
A short computation shows that our formula \eqref{datyp} for the
characters of discrete series representations provides a function
$\chi_{[k/4-1/2]_-}$ with the desired property. This constitutes a
rather strong test for the expression we proposed.

\subsubsection{The continuous series representations}

As we have discussed above, the Lie superalgebra $\psl$ possesses
another type of infinite dimensional irreducible representations,
the principal continuous series. These representations are
labelled by pairs $\bigl[(j_1,\alpha),j_2\bigr]^\wedge$ where $j_1 \in
\mathbb{S}=1/2+i\Real$ and $0\leq\alpha<1$. These representations
give rise to typical representations of the affine Lie superalgebra.
Their characters are,
\begin{align}
\chi_{[(j_1,\alpha),j_2]}(q,z_1,z_2) &\ =\ \frac{1}{\eta^{4}(q)}
  \, \prod_{\nu,\mu = \pm} \, \vartheta_1(z_1^{\nu /2}
 z_2^{\mu /2},q) \ \chi^{-k-2}_{(j_1,\alpha)}(q,z_1)\ \chi^{k-2}_{j_2}
 (q,z_2)\\[2mm]
 \text{where} \ \ \chi^{-k-2}_{(j_1,\alpha)}(q,z_1) &\ =\ %
   -i q^{-\frac{(j_1-1/2)^2}{k}}
   \, \chi_{(j,\alpha)}(z_1) \, (z_1^{1/2}-z_1^{-1/2}) \, \vartheta_1(q,z_1)^{-1}\ \ .
\end{align}
These characters will be the most important building blocks for
the partition sum of the bulk $\PSU$ WZNW model.

\subsection{Representations with a finite number of ground states}

Finally, we turn to representations with finite number of ground
states. These do not appear in the bulk spectrum of the $\PSU$ WZNW
model but are expected to furnish the building block for the boundary
spectra associated with instantonic branes. Special attention is
devoted to the atypical representations.

\subsubsection{Typical representations}

The free field construction we have reviewed above suggests that
generic representations have no singular vectors except from the
ones that arise through the representations of the two bosonic
$\asl$ current algebra at levels $\pm k-2$. This is indeed the
case. The statement implies a precise expression for the
characters of typical representations
\begin{equation} \label{charfintyp}
\chi_{[j_1,j_2]}(q,z_1,z_2) \ = \ \frac{1}{\eta^{4}(q)}
  \, \prod_{\nu,\mu = \pm} \, \vartheta_1(q,z_1^{\nu /2} z_2^{\mu /2})
 \ \chi^{-k-2}_{j_1}(q,z_1)\,    \chi^{k-2}_{j_2} (q,z_2)
\end{equation}
where $j_1 \neq j_2$ and $j_1 \leq k/2+1$. The function $\vartheta_1$ was
defined in eq.\ \eqref{chargentyp} above. We also recall that the
$\asl$ characters for negative level are given by
\begin{equation}
   \chi^{-k-2}_{j_1} (q,z_1) \ = \ i q^{-\frac{(j_1+1/2)^2}{k}}
   \, \left(z_1^{j_1+1/2} - z_1^{-j_1-1/2}\right) \,
   \vartheta_1(q,z_1)^{-1}
\end{equation}
We shall use the symbol $[j_1,j_2]^\wedge$ for these irreducible
representations of the affine algebra.

\subsubsection{Atypical representations}

Nothing prevents us from evaluating the previous character
formulas at the points $j_1 = j = j_2$. But the resulting
functions turn out to be the characters of indecomposable
representations $[j,j]^\wedge$ which contain fermionic singular
multiplets. The latter lie all within the space of ground states
and hence the decomposition follows exactly the decomposition
formulas (\ref{Kj},\ref{K0},\ref{K12}) for Kac modules of the Lie
superalgebra $\psl$
\begin{equation}\label{aatyp}
\begin{split}
\chi_{[j,j]}(q,z_1,z_2) & = \
   2\chi_{[j]}(q,z_1,z_2)
  - \chi_{[j+1/2]}(q,z_1,z_2)
  - \chi_{[j-1/2]}(q,z_1,z_2) \\[2mm]
\chi_{[1/2,1/2]}(q,z_1,z_2) & =\
   2\chi_{[1/2]}(q,z_1,z_2)
  - \chi_{[1]}(q,z_1,z_2)
  - 2\chi_{[0]}(q,z_1,z_2)\\[2mm]
\chi_{[0,0]}(q,z_1,z_2) & = \
   2\chi_{[0]}(q,z_1,z_2)
  - \chi_{[1/2]}(q,z_1,z_2)\ \ .
\end{split}
\end{equation}
In the first row we assumed $j \geq 1$. The relative sign between
the two terms on the right hand side is due to the fermionic
nature of the singular vectors. Even though equation \eqref{aatyp}
is not a closed formula for the characters of atypical
representations, it can be used to construct the latter
recursively as a sum of the functions $\chi_{[j,j]}$
and of the vacuum character $\chi_{[0]}$.
We have determined the latter in the previous
subsection and hence know all characters, at least implicitly.
\smallskip

We can do a little better, though, and provide explicit
expressions for the atypical characters. To this end we observe
that the typical representations obey
\begin{equation*}
 \chi_{[j_1,j_2]}(q,z_1,z_2)
  \ = \ \chi_{[(-,j_1+1),j_2]}(q,z_1,z_2)
        - \chi_{[(+,j_1+1),j_2]}(q,z_1,z_2) \ \ ,
\end{equation*}
where we silently agreed to formally expand {\em both}
characters on the right hand side in powers of $z_1$. Using this
observation, we deduce that the characters of atypical
representations $[j]^\wedge, j \geq 1/2,$ must be given by
\begin{align}\nonumber
\chi_{[j]}(q,z_1,z_2) &\ =\ -\sum_{n=0}^\infty \
   (n+1) \, \chi_{[j+1/2 + n/2,j+1/2+n/2]}(q,z_1,z_2)\\[2mm]
&\ =\ \chi_{[j]_+}(q,z_1,z_2)-\chi_{[j]_-}(q,z_1,z_2) \ \ \
\text{ for } \ \ j \geq \frac12\ . \label{charfinatyp}
\end{align}
It is somewhat cumbersome but
rather straightforward to check that these characters indeed obey
the relations \eqref{aatyp}. Note that formula \eqref{charfinatyp}
should only be used for $j \geq 1/2$. The vacuum character is not
given by expression \eqref{charfinatyp} but rather by formula
\eqref{vacchar}.

\newpage
\noindent
{\huge{\bf Part II: Solution of the WZNW model} }\\[2cm]
After all the representation theoretic preparations we can now
address the WZNW model on the supergroup $\PSU$. We shall start
with a discussion of the Lagrangian before we proceed to the
particle limit and analyze its state space in some detail. We
shall show, in particular, that the Laplacian on the supergroup
$\PSU$ is non-diagonalizable and obtain explicit formulas for all
its generalized eigenfunctions. Afterwards we turn to the full
field theory. The latter will be solved through a free field
representation. We discuss how the non-diagonalizability of the
Laplacian is naturally inherited by the zero-mode of the energy
momentum tensor. Hence, the $\PSU$ WZNW model provides an example
of a logarithmic conformal field theory. As an application of our
results on the structure of the state space, we finally propose an
algorithm that allows to count the number of states that possess
the same transformation law under the global symmetries. The
resulting formulas will only be used in our forthcoming analysis
of the RR deformation. Nevertheless, some comments on the latter
are included at the end of this part.

\section{The WZNW Lagrangian}

Before we spell out the WZNW model we are about to consider, we would like to
recall a few basic facts on supergroups. Let us begin with the supergroup
$\GLmn$. Elements $S\in\GLmn$ can be represented through
invertible matrices of the form
\begin{equation} \label{SOne}
  S\ = \ \mat A&\sigma\\\rho&B\tam
\end{equation}
where $A$ and $B$ are elements of $\text{GL}(m)$ and $\text{GL}(n)$, respectively, with
Grassmann-even matrix elements. The symbols $\sigma$ and $\rho$, on the
other hand, denote rectangular matrices with Grassmann-odd entries. We
pass from $\GLmn$ to $\SLmn$ by imposing the additional condition
$\sdet(S)=1$ on the superdeterminant of $S$,
\begin{equation}
\sdet(S)\ =\ \frac{\det(A-\sigma B^{-1}\rho)}{\det(B)}\ =\
 \frac{\det(A)}{\det(B-\rho A^{-1}\sigma)}\ \stackrel{!}{=}\ 1 \ \ .
\end{equation}
When $n=m$, the construction of the superdeterminant implies that SL(n$|$n)
possesses a non-trivial center consisting of scalar multiples of the
identity matrix. In descending to $\PSLnn$, we identify supermatrices
in $\SLnn$ that differ by a scalar multiple. Furthermore, we
introduce the following unitarity condition for supermatrices $S\in\GLmn$,
\begin{equation}  S \, \eta \, S^\dagger \ = \ \eta
\label{ad}
\end{equation}
where $\eta = {\text{diag}} (-1,1,\dots ,1)$ is the $m+n$ dimensional
Minkowski metric. Supermatrices $S \in\SLmn$ obeying the condition
\eqref{ad} form the supergroup $\text{SU}(1,m-1|n)$. Identification of scalar
multiples finally leads to $\text{PSU}(1,m-1|n)$.
\smallskip

We are now prepared to spell out the action functional for the $\WZNW$ model
on $\PSU$ at level $k$. Since there is no fundamental representation
of this group we will instead work with $\text{SU}(1,1|2)$ and show that the Lagrangian
actually does not depend on the additional degree of freedom corresponding
to multiples of the identity element. For all $S\in\text{SU}(1,1|2)$ we thus define
\begin{equation}
  \label{eq:WZW}
  \mc{S}_k^{\text{SU}(1,1|2)}[S]\ =\ -\frac{k}{2\pi}\int\!d^2z \;\str  \left(S^{-1}\partial S S^{-1}
  \bar{\partial}S \right) -\frac{k}{12\pi} \int\;\str
  \left(\left(S^{-1}dS \right)^3 \right)
\end{equation}
with a suitably normalized supertrace $\str$. For supermatrices $S$ of the
form \eqref{SOne}, the supertrace is given by $\str(S)=\tr(A)-\tr(D)$. Using the
Polyakov-Wiegmann identity for WZNW models,
\begin{equation} \label{PWI}
\mc{S}_k^{\text{SU}(1,1|2)}[S_1S_2]\ =\ \mc{S}_k^{\text{SU}(1,1|2)}[S_1]
+\mc{S}_k^{\text{SU}(1,1|2)}[S_2] +\frac{k}{2 \pi} \int\!d^2z\; \str
\left(S_1^{-1}\bar{\partial}S_1 \partial S_2 S_2^{-1} \right)\ \ ,
\end{equation}
one may easily show that the action $\mc{S}_k^{\text{SU}(1,1|2)}$
remains invariant if we multiply the supermatrix $S\in\text{SU}(1,1|2)$
with a scalar factor $\exp \Phi$, i.e.\ %
$$ \mc{S}_k^{\text{SU}(1,1|2)}[e^\Phi S] \ = \ \mc{S}_k^{\text{SU}(1,1|2)}[S] \ \ .$$
This relation ensures that the functional form of the WZNW action for $\PSU$ is
identical to the WZNW action for $\text{SU}(1,1|2)$. In particular we like to stress
that no explicit gauge procedure is required, in contrast to what has been proposed in
\cite{Bhaseen:1999nm}. As in all WZNW models the Lagrangian defined in
\eqref{eq:WZW} leads to two chiral sets of currents $J^\mu(z)$ and $\bar{J}^\mu(\bar{z})$
which generate two (anti)commuting copies of the affine Lie superalgebra
$\apsu(1,1|2)_k$. Their precise relations can easily be reconstructed
from its zero-mode subsuperalgebra \eqref{eq:CR} and the metric \eqref{eq:Metric}.
\smallskip

Our aim is to show that the introduction of auxiliary fields
allows to decouple bosonic and fermionic degrees of freedom to a
large extent. The result will be a sum of two renormalized bosonic
WZNW models, the action for a holomorphic and anti-holomorphic set
of symplectic fermions and an interaction term which mixes the
two. Our derivation is motivated by the ideas given
in~\cite{Bhaseen:1999nm} but the fermions are treated differently,
along the lines of \cite{Schomerus:2005bf}. The first step is to
rewrite our action with the help of the following product
decomposition of $\text{SU}(1,1|2)$ supermatrices,
\begin{equation}\label{STwo}
  S\ =\ e^{\Phi}\mat\id&0\\\lambda&\id\tam\cdot\mat A&0\\0&B
    \tam\cdot\mat\id&\chi\\
    0&\id\tam\ =\ e^{\Phi}\mat A&A\chi\\\lambda
      A&\lambda A\chi+B\tam \ \
\end{equation}
where the factor $\exp \Phi$ is chosen such that the matrices $A$ and
$B$ are uni-modular. Using once more the Polyakov-Wiegmann identity
\eqref{PWI} for supergroups, we find that
\begin{align}
  \mc{S}_k^{\PSU}[S] &\ =\ %
      \mc{S}_k^{\PSU}\left[\mat\id&0\\\lambda&\id\tam\right]+
      \mc{S}_k^{\PSU}\left[\mat A&0\\0&B\tam\right] +
      \mc{S}_k^{\PSU}\left[\mat\id&\chi\\0&\id\tam\ \right] \nonumber \\[3mm]
      & \hspace*{3cm}
    + \frac{k}{2\pi}\int\!d^2z\;\str \mat 0&0\\\bartial\lambda\partial A
      A^{-1}& \bartial \lambda A  \partial \chi
            B^{-1}\tam
   \label{lagrange}
\end{align}
The first and third term vanishes due to the fact that the only contributions
to supertraces come from non-trivial bosonic submatrices. Hence, we are left
with only two terms. Taking into account the indefinite structure of the metric
the result can now we rewritten as
\begin{align}
  \mc{S}_k^{\PSU}[S]\ = \ \mc{S}_k^{\AdS_3}[A] + \mc{S}_k^{\SU}[B]
- \frac{k}{2\pi}\int\!d^2z\,\tr\Bigl[\bartial \lambda A \partial \chi B^{-1}\Bigr]
\label{finalaction}
\end{align}
where $\mc{S}_k^{\AdS_3}$ and $\mc{S}_k^{\SU}$ denote the usual bosonic
WZNW actions. It is easy to see that the previous action is equivalent to
  the functional
\begin{equation}
  \label{eq:DecoLagrangian}
  \mc{S}_k^{\PSU}[S,p,\theta]\ =\ \mc{S}_{k+2}^{\AdS_3}[A]\ +
    \ \mc{S}_{k-2}^{\SU}[B]\ + \frac{1}{2\pi k}\int\!d^2z\,
      \tr\bigl\{k(p\bartial\theta+\bar{p}\partial\bar{\theta})
      +A^{-1}pB\bar{p}\bigr\}
\end{equation}
  where the fields have been decoupled to a large extent. Integrating
  out the auxiliary fields $p$ and $\bar{p}$ one indeed arrives
  at the original action if one imposes the identifications $\theta=\lambda$
  and $\bar{\theta}=\chi$. The shift of the levels arises from
  the modification of the path integral measure.

\section{\label{sc:MS}The minisuperspace theory}

In this section we would like to analyse the space of
(generalized) eigenfunctions of the Laplace operator on the
supergroup $\PSU$. We shall set up the problem in the first
subsection. Explicit formulas for all the generalized
eigenfunctions are derived in the second subsection. Their
transformation behaviour with respect to the left and right
regular action of $\psl$ is finally investigated in the
last subsection.

\subsection{The Laplacian on the Supergroup $\mathbf{PSU(1,1|2)}$}

On the supergroup $\PSU$ we can introduce various different
coordinates. For our purposes in the next subsection, a
preferred set of coordinates is defined through the
decomposition,
\begin{equation}
  G\ =\ e^{i\eta_a S_2^a}\,g\,e^{i\bar{\eta}_a S_1^a}
   \ =\ e^{i\eta_a S_2^a}\,e^{ix_{ab} K^{ab}}\,e^{i\bar{\eta}_a S_1^a}\ \ .
\end{equation}
In these coordinates, we can easily express the differential
operators
implementing the left and right regular representation. In the following, we
denote the generators of the bosonic subalgebra on the bosonic
subgroup by $L^{ab}_0$ and $R^{ab}_0$, respectively. For the left
regular action we find (see also \cite{Dolan:1999dc})
\begin{align}
L_2^a \ = \ \pl^a \ \ \ \  & , \ \  \ \ L^{ab} \ = \
L^{ab}_0 - i (\eta^a\pl^b - \eta^b \pl^a) \\[2mm]
L_1^a \ = \ -D^{ab}(g)\,\bar\pl^b + \tilde L_1^a \ \ \ \ &
\text{where} \ \ \tilde L_1^a \ = \ \frac{i}{2}\,\e^{abcd}\,
\left(\eta^b L^{cd}_0 - i \eta^b\eta^c \pl^d \right)\ \ .
\label{tildeRp}
\end{align}
Here, the partial derivatives $\pl^a$ and $\bar \pl^a$ denote
differentiation with respect to the fermionic coordinates $\eta_a$
and $\bar \eta_a$, respectively, and the matrix $D^{ab}(g)$ of
functions on the bosonic subgroup is obtained by evaluation of
elements $g$ in the $(1/2,1/2)$ representation. By construction,
it satisfies the following differential equations
\begin{equation*}
  L^{ab}_0 \, D^{cd}(g) \ = \ i \bigl(\d^{ac} D^{bd}(g) -
    \d^{bc} D^{ad}(g)\bigr)  \ \ , \ \
   R^{ab}_0 \, D^{cd}(g) \ = \ i \bigl(\d^{ad} D^{cb}(g) -
    \d^{bd} D^{ca}(g)\bigr)\ \ .
\end{equation*}
To check the commutation relations is straightforward, though a bit
cumbersome. The right regular representation is obtained similarly,
with the two types of fermionic generators exchanged, $\bar \eta_a$
replacing $\eta_a$ etc.\  Needless to stress that the left and right
action \mbox{(anti-)commute}.
\smallskip

A short and straightforward computation of the quadratic Casimir
element
$C_2=-\epsilon_{abcd}\,K^{ab}K^{cd}/4+\epsilon_{\alpha\beta}S_\alpha^aS_\beta^a$
in the left or right regular representation gives
the following explicit formula for the Laplacian $\Delta$ on the
supergroup
\begin{equation}
\Delta \ = \ L(C_2)\ = \ R(C_2)\ = \Delta_0 + Q \ \ \
\text{ where }\ \ \
Q \ = \ 2\,\partial_a D^{ab}(g) \bartial_b\
\end{equation}
and $\Delta_0$ is the usual Laplace operator on the bosonic
subgroup, i.e.\ on $\AdS_3\times\text{S}^3$. Our aim is to
find generalized eigenfunctions of this
operator. Let us recall that a function $\psi$ is called a
generalized eigenfunction of $\Delta$ for eigenvalue $\lambda$ if
\begin{equation}\label{gef}
(\Delta - \lambda)^n \, \psi \ = \ 0 \ \ \ \ \
 \text{ for some } \ \ \ \ n > 0 \ \ .
 \end{equation}
$\psi$ is an eigenfunction if this equation is satisfied for $n=1$
(and hence for all other values of $n$). We shall see in the next
subsection that generalized eigenfunctions of $\Delta$ with
$\lambda \neq 0$ are in fact true eigenfunctions. For $\lambda =
0$, on the other hand, non-trivial generalized eigenfunctions do
appear. This means that the Laplacian on the supergroup $\PSU$ can
only be brought into Jordan normal form. The Jordan cells turn
out to possess a rank up to five. By construction, the
individual spaces
of generalized eigenfunctions come equipped with the left and
right regular action of the Lie superalgebra $\psl$. We shall
describe its decomposition into indecomposables in the second
subsection.

\subsection{Generalized eigenfunctions of the Laplacian}

In this subsection we show that generalized
eigenfunctions of the Laplacian on $\PSU$ are in one-to-one
correspondence with elements of the following auxiliary
space
\begin{equation}
  \label{eq:AuxSpaceOne}
\mc{H}_0 \ := \ L_2\bigl(\text{AdS}_3 \times \SU\bigr) \otimes \Lambda
(\eta,\bar \eta)
\end{equation}
of Grassmann valued functions whose coefficients are square
integrable functions on the bosonic subgroup. Under favorable
circumstances, i.e.\ when the fermionic generators of the
Lie superalgebra transform in a unitary representation of
its bosonic subgroup $G$, the space of Grassmann valued
functions on $G$ coincides with the space of generalized
eigenfunctions. This is the case e.g.\ %
for the $\GL$ model studied in \cite{Schomerus:2005bf}, but
the key example we have in mind here is $\PSL$ with a real
form that removes all non-trivial finite dimensional
representations of the bosonic subalgebra, and in particular
the $(1/2,1/2)$ representation, from the list of unitaries.
\smallskip

The key idea in the subsequent construction of generalized
eigenfunctions is to consider the elements $\psi_0$ of
${\cal H}_0$ as `leading contributions'. More precisely,
we shall show that all eigenfunctions of the bosonic
Laplacian can be turned into
generalized eigenfunctions by adding appropriate terms
with lower fermion number. To this end, let us rewrite
equation \eqref{gef} in the following form
$$ (\Delta - \lambda)^n \psi_\lambda \ = \ \bigl((\Delta_0-\lambda)^n +
A_n(\lambda)\bigr) \psi_\lambda \ = \ 0 \ \ . $$
The operators $A_n(\lambda)$ are lengthy combinations of $Q$ and
$(\Delta_0-\lambda)$ which can be worked out explicitly.
What is most important is to note that each term in $A_n(\lambda)$
contains a least one $Q$. Hence, the operators $A_n(\lambda)$ are
nilpotent.
\smallskip

A short and formal manipulation shows that generalized
eigenfunctions for the eigenvalue $\lambda$ at order $n$
possess the general form
$$ \psi_\lambda^{(n)} \ = \ \Xi_\lambda^{(n)} \psin \ = \ \sum_{\nu=0}^\infty
\bigl( - (\Delta_0-\lambda)^{-n} A_n(\lambda)\bigr)^\nu \psin $$
where $\psin \in {\cal H}_0$ is an eigenfunction of $\Delta_0$
with eigenvalue $\lambda$, i.e.\ $\Delta_0 \psin = \lambda \psin$.
Since $A_n(\lambda)$ contains anti-commuting fermionic
derivatives, the sum on the right hand side truncates after a
finite number of terms at $\nu = 4$. On the other hand, the
formula requires to invert the operator $\Delta_0-\lambda$. Hence
the operator $\Xi^{(n)}_\lambda$ need not be well defined for all
$\psin$. To analyze this issue further, we note that
$\Xi_\lambda^{(n)} = \Xi_\lambda^{(5)} = \Xi_\lambda$ is
independent of $n$ for $n \geq 5$ and compute
\begin{align}\nonumber
\psi_\lambda\ =\
\Xi_\lambda \psin &\ =\ %
 \psin - \frac{1}{(\Delta_0-\lambda)}Q'_\lambda \psin +
\left(\frac{1}{(\Delta_0-\lambda)}Q'_\lambda\right)^2
\, \psin + \frac{1}{(\Delta_0-\lambda)^2} Q'_\lambda
Q''_\lambda \psin + \dots \\[4mm]
& \text{ where } \ \  Q \ = \ Q'_\lambda + Q''_\lambda \ = \
\bigl(1-P_0(\lambda)\bigr) Q + P_0(\lambda) Q
\label{expgef}
\end{align}
and $P_0(\lambda)$ is the projector on eigenstates of $\Delta_0$
with eigenvalue $\lambda$. We have not displayed the third and
fourth order terms in $Q$ because the expression would be rather bulky
(there are 14 such terms). Let us observe that the inverse powers of
$(\Delta_0 - \lambda)$ only act in combination with $Q'_\lambda$,
i.e.\ after application of the projection $1-P_0(\lambda)$. This
continues to hold for the higher order terms and hence $\psi_\lambda$ is
well defined for all $\psin$. Since $P_0(\lambda) \Xi_\lambda =
\text{id}$, we conclude that generalized eigenfunctions are indeed in
one-to-one correspondence with elements $\psi_0 \in {\cal H}_0$.
\smallskip

It is instructive to contrast these findings with results on the
true eigenvectors of the Laplacian. Our general formula
\eqref{gef} applied to the special case $n=1$ shows that such
eigenfunctions must be of the form
\begin{align} \label{ef}
\psi_\lambda^{(1)} &\ =\ \Xi_\lambda^{(1)} \psin \ = \
 \sum_{\nu=0}^\infty \left(-(\Delta_0 - \lambda)^{-1} Q\right)^\nu
 \psin\\[2mm] \nonumber
 &\ =\ \psin - \frac{1}{(\Delta_0-\lambda)}Q \psin +
\left(\frac{1}{(\Delta_0-\lambda)}Q\right)^2 \, \psin + \dots
\end{align}
In order for $\psi_\lambda^{(1)}$ to be well defined it is obviously necessary
that $Q''_\lambda \psin = P_0(\lambda) Q \psin = 0$.\footnote{The
condition is also sufficient, though this requires a slightly
more elaborate argument.} Note that the condition automatically
ensures that our expression for the eigenfunction $\psi_\lambda^{(1)}$ agrees
with the formula for generalized eigenfunctions above.
It is easy to see that
\begin{equation} \label{claim}
 P_0(\lambda) Q \psin \ = \ 0 \ \ \ \text{ for eigenfunctions
$\psin$ of $\Delta_0$ with eigenvalue $\lambda \neq 0$}\ \ .
\end{equation}
Hence, we conclude that all generalized eigenfunctions of the
Laplacian on $\PSU$ with nonzero eigenvalue are true eigenfunctions.
For $\lambda=0$, on the other hand, non-trivial generalized
eigenfunctions exist. These are in one-to-one correspondence
with functions $\psi_0^\lambda$ for which
$P_0(0)Q \psi_0^\lambda \neq 0$.
\smallskip

Our claim \eqref{claim} can be established as follows: suppose
that $\psin$ transforms in the representation $(j_1,j_2)$ of the
bosonic subalgebra. Then, after application of $Q$, the resulting
state $Q\psin$ decomposes into four composents according to the
four different representations $(j_1\pm1/2, j_2 \pm 1/2)$ and
$(j_1\pm 1/2,j_2\mp1/2)$ that arise after multiplication with the
functions $D^{ab}(g)$ in the $(1/2,1/2)$ representations. The
resulting possible eigenvalues of the bosonic Laplacian are
$\delta_\pm = \lambda \pm (j_1-j_2)$  and $\delta_\pm = \lambda
\pm (j_1 + j_2 - 1)$ with $\lambda = j_1(j_1+1) - j_2 (j_2+1)$.
Hence, we conclude that $P_0(\lambda) Q \psin = 0$, unless $j_1 =
j_2$ or $j_1+1=-j_2$. The latter conditions on the choice of
$(j_1,j_2)$ are equivalent to requiring $\lambda = 0$. This proves
our claim and concludes this subsection.

\subsection{Regular action on generalized eigenfunctions}

Since the Laplacian commutes with both the left and the right
regular representation, $L$ and $R$ provide two (anti-)commuting
actions of the Lie superalgebra $\psl$ on generalized
eigenfunctions $\psi \in {\cal H} \equiv \Xi {\cal H}_0$.
\smallskip

In order to spell out the behavior of states $\psi \in {\cal H}$
under the right regular action we introduce the following new
representations
$$ B(\mu,\nu) \ :=\  {\rm Ind}_{\g^{(0)}}^{\g}
  V_{(\mu,\nu)} \ = \ \mc{U}(\g) \otimes_{\g^{(0)}} V_{(\mu,\nu)} \
\ .
$$
By construction, these representations have a dimension $16^2
\cdot\text{dim}(\mu,\nu)$ and certainly none of them is irreducible.
All the representations can be decomposed into a sum of projective
representations \cite{MR1378540}. Generically, $B(\mu,\nu)$
decomposes into a sum of typical Kac modules according to
\begin{equation}
B(\mu,\nu) \ \cong \bigoplus_{(\mu',\nu')} \ \bigl[ [\mu,\nu] :
 (\mu',\nu')\bigr]\, \cdot\, [\mu',\nu']\label{Bdec1}
 \end{equation}
where $\bigl[[\mu,\nu] :(\mu',\nu')\bigr]$ denotes the multiplicity of the
bosonic multiplet $(\mu',\nu')$ inside the Kac module $[\mu,\nu]$.
The formula applies whenever the summation extends only over
bosonic representations $(\mu',\nu')$ with non-vanishing Casimir.
More explicitly, we can use the above formula for all
representations $(\mu,\nu) = \bigl((j_1,\a),j_2\bigr)$ from the continuous
series and for discrete series representations $(\mu,\nu) =
\bigl((\pm,j_1),j_2\bigr)$ as long as $j_1+j_2 + 1 \neq 0, \pm 1$. In the
remaining cases, projective covers of atypical representations appear,
\begin{align} \label{Bdec2}
B\bigl((\pm,-j-1),j\bigr) &\ \cong\   2 \cdot \P_j^\pm
  \ \oplus \ \dots \\[2mm] \label{Bdec3}
B\bigl((\pm,-j - \foh),j + \foh\bigr) &\ \cong\  \P_{j-\foh}^\pm \
\oplus \ \dots \\[2mm] \label{Bdec4}
B\bigl((\pm,-j - \fth),j - \foh\bigr) &\ \cong\  \P_{j+\foh}^\pm\ \oplus \
\dots \ \
  \end{align}
where the dots $\dots$ stand for a sum of typical Kac modules that
can be determined through the rule \eqref{Bdec1} if we remember to
omit all terms that correspond to an atypical representation.
\smallskip

It is relatively easy to see that the space ${\cal H}$ of
generalized eigenfunctions possesses the following decomposition
with respect to the asymmetric action of the subsymmetry
$\g^{(0)}_L \oplus \g_R$,
\begin{align} \nonumber
{\cal H} &\ \cong\ %
\bigoplus_{2J=0}^\infty\int_{\mathbb{S}}dj\int_0^1d\alpha\
\bigl((j,\alpha),J\bigr)^+_L\,  \otimes\,  B\bigl((j,\alpha),J\bigr)_R\ \oplus \\[2mm]
 &\ \hspace*{2cm} \oplus \ \bigoplus_{2J=0}^\infty\bigoplus_{\eta=\pm}
  \int_{\foh}^\infty dj \ \bigl((\eta,-j),J\bigr)^+_L \, \otimes\,  B\bigl((\eta,-j),J\bigr)_R\ \
  \ \ .\nonumber
\end{align}
The domain of the first integral is given by $\mathbb{S} = -1/2 +
i \mathbb{R}$, as usual for the principal continuous series
representations. In order to justify the decomposition, let us
note that the ground states from which the representations
$B(\mu,\nu)$ are generated, can be identified with those states
$\psi_\lambda$ in ${\cal H}$ whose top component $P_0(\lambda)\psi_\lambda$ has
maximal fermion number.
\smallskip

In order to rewrite our decomposition formula in terms of
indecomposables, we need to insert the formulas
\eqref{Bdec1}-\eqref{Bdec4}. We then collect all terms that give
rise to the same projective representation of $\g_R$. On the
subspace $\Htyp\subset\mc{H}$ of states with non-zero eigenvalue $\lambda$,
the bosonic multiplets of of the $\g^{(0)}_L$ action turn out to
combine into a Kac module for the Lie superalgebra $\g_L$, i.e.
\begin{align*}
\Htyp &\ \cong\ %
\bigoplus_{2J=0}^\infty\int_{\mathbb{S}}dj\int_0^1d\alpha\
\bigl[(j,\alpha),J\bigr]^+_L\,  \otimes\,  \bigl[(j,\alpha),J\bigr]_R\ \oplus \\[2mm]
 & \hspace*{2cm} \oplus \ \bigoplus_{2J=0}^\infty\bigoplus_{\eta=\pm}
  \int_{\foh,j\neq J+1}^\infty dj \ \bigl[(\eta,-j),J\bigr]^+_L \, \otimes\,  \bigl[(\eta,-j),J\bigr]_R\ \
  \ \ .
\end{align*}
Another way to arrive at this result is by noting that our
operator $\Xi$ provides an intertwiner between the action of $L$
and $R$ on $\Htyp$ and some simplified action $\tilde L$ and
$\tilde R$ of $\psl$ on the subspace $\mc{H}_{0,\text{typ}}\subset\mc{H}_0$ of
eigenfunctions with non-zero eigenvalues of $\Delta_0$. The action
$\tilde L$, for example, is generated by the operator $\tilde
L^a_1$ defined in formula \eqref{tildeRp} along with
\begin{equation}
  \label{tildeRpMore}
\tilde L^a_2 \ = \ L^a_2 \ \ \ \ , \ \ \ \tilde L^{ab} \ = \
L^{ab} \ \ .
\end{equation}
A similar construction gives $\tilde R$. With respect to these two
actions of $\psl$, the space ${\cal H}_0$ is easily seen to
decompose into an integral over left and right Kac modules. This
applies even to the subspace on which $\Delta_0$ vanishes. But on
the latter $\Xi$ ceases to be an intertwiner between the truncated
actions $\tilde L, \tilde R$ and the full regular action $L,R$. In
the next section we shall see that the $\tilde L - \tilde R$
module ${\cal H}_0$ models the space of vertex operators in the
free field representation. The full $L-R$ action on ${\cal H}$, on
the other hand, agrees with actions of $\psl$ on the ground states
of the full interacting theory. The discrepancy between the two
actions in the atypical sector will have remarkable consequences
which at the end culminate in the logarithmic behavior
of the WZNW theory.
\smallskip

As for the atypical sector $\Hatyp$ of generalized
eigenfunctions with vanishing eigenvalue $\lambda$, it is built up
from projective covers only when considered with respect to the
right (or left) regular action. The associated multiplicity spaces
possess the same $\g^{(0)}_L$ representation content as the
atypical irreducible $\psl$ representations from the discrete series. But this
time, enhancing the left action from the bosonic subalgebra to the
full $\psl$ has more drastic effects than simply to promote the
multiplicity spaces into representations of the Lie superalgebra.
Note that such a behavior would obviously violate the symmetry
between left and right regular transformations and hence cannot be
the right answer. Instead, as a $\g_L \oplus \g_R$ module, the
atypical sector ${\cal H}$ is built from non-chiral indecomposables
which encompass an infinite number of atypical constituents much
in the same way as it happens for $\GL$ (see \cite{Schomerus:2005bf}).
We refrain from working out the details here.
\smallskip

This gives us a fairly complete picture of the space of wave
functions for a particle moving on $\PSU$ and a very good basis to
discuss how field theoretic corrections affect the structure of
the state space. In the full field theory, there will be two new
phenomena which have to be taken into account. First of all there
will be a cut-off associated with the finiteness of the level $k$.
Moreover, the affine Lie superalgebra admits a family of spectral
flow automorphisms which has to be taken into account properly.

\section{Vertex operators and correlation functions}

  Now that we obtained a profound knowledge about the particle limit
  of the sigma model on $\PSU$, we are finally in a position to
  return to the solution of the full quantum theory. Our starting point
  is the Lagrangian \eqref{eq:DecoLagrangian},
\begin{equation}
  \label{eq:WZWnew}
  \mc{S}_k^{\PSU}
  \ =\ \mc{S}_0+\mc{S}_{\text{int}}
  \ =\ \mc{S}_{k+2}^{\AdS_3}+\mc{S}_{k-2}^{\SU}+\mc{S}_{\text{ferm}}+\mc{S}_{\text{int}}\ \ ,
\end{equation}
  consisting of a decoupled system with a purely bosonic WZNW model on
  $\AdS_3\times\SU$ and a set of free fermions as well as an interaction
  term coupling bosonic and fermionic degrees of freedom. Following the
  general strategy adopted in \cite{Schomerus:2005bf} we will start
  with an analysis of the decoupled theory and consider
  the additional term as a perturbation. Our main aim is to find
  the vertex operators of the full supergroup WZNW theory and to sketch
  the calculation of their correlation functions.
\smallskip

  The state space $\mc{\hat{H}}_0$ of the decoupled theory described by
  $\mc{S}_0$ is completely known using standard results in conformal
  field theory. For reasons to become clear below we restrict ourselves
  to fermions with integer moding. Under this assumption there exists a
  unique representation $\mc{F}\otimes\mc{\bar{F}}$ for the fermions.
  It is generated from a ground state by the application of the modes
  $\theta_{-n}^a$ and $p_{-(n+1)}^a$ for $n\geq0$ and similarly for the
  anti-holomorphic fields. The $\SU_{k-2}$ WZNW model is described by
  a charge conjugate partition function involving unitary representations
  with spin $2J=0,1,\dots,k-2$ \cite{Gepner:1986wi}. In the $\AdS_3$
  WZNW model on the other hand two different kinds of representations
  of the underlying affine algebra $\widehat{\text{sl}}(2,\Real)_{k+2}$
  contribute \cite{Maldacena:2000hw}: the principal continuous series
  $(j,\alpha)$ for $j\in\mathbb{S}=-\frac{1}{2}+i\Real$ and $\alpha\in[0,1)$ and the
  discrete series $(\pm,j)$ for $-\frac{1}{2}>j>-\frac{k+1}{2}$.
  Moreover, one has to take into account the spectral flow automorphism
  which maps ordinary highest weight modules to twisted ones. This leads to
  an additional quantum number $w$ which has to be attached to the representations
  of $\widehat{\text{sl}}(2,\Real)_{k+2}$. The precise definition of the spectral
  flow automorphism has been given in \eqref{eq:SF}.
\smallskip

Based on the previous remarks we can spell out the space of the
decoupled system,\footnote{We remind
  the reader that the orbit of representations $(+,j)_w$ includes
  representations based on $(-,j')$.}
\begin{equation}
  \label{eq:StateDeco}
  \begin{split}
    \mc{\hat{H}}_0
    &\ \cong \ \bigoplus_{2J=0}^{k-2}\bigoplus_{w\in\Integer}\int_{\mathbb{S}}dj\int_0^1\!\!\!d\alpha\,
\Bigl[\mc{V}_{((j,\alpha)_w,J)}\otimes\mc{F}\Bigr]\otimes\Bigl[\mc{V}_{((j,\alpha)_w,J)^+}\otimes\mc{\bar{F}}\Bigr]\\[2mm]
    &\qquad\qquad\oplus\bigoplus_{2J=0}^{k-2}\bigoplus_{w\in\Integer}
  \int_{\frac{1}{2}}^{\frac{k+1}{2}}\!\!\!dj\,\Bigl[\mc{V}_{((+,-j)_w,J)}\otimes\mc{F}\Bigr]\otimes\Bigl[\mc{V}_{((+,-j)_w,J)^+}\otimes\mc{\bar{F}}\Bigr]\ \ .
  \end{split}
\end{equation}
  Nevertheless we are not yet done. Since we intend to describe a
  supersymmetric theory we have to cast the state space in a
  manifestly covariant form. In addition we have to make
  contact to the minisuperspace analysis presented in
  section \ref{sc:MS}. In order to achieve these goals
  we must find a realization of each, the holomorphic and anti-holomorphic
  affine Lie superalgebras $\widehat{\text{psu}}(1,1|2)_k$, on $\mc{\hat{H}}_0$. Moreover
  their zero mode action on the ground states has to resemble that of the
  differential operators $\tilde{L}$ and $\tilde{R}$ on $\mc{H}_0$,
  respectively. This space and the corresponding operators have been
  introduced in \eqref{eq:AuxSpaceOne} and \eqref{tildeRpMore}.
\smallskip

  In fact, a realization of the affine Lie superalgebra $\widehat{\text{psu}}(1,1|2)_k$
  in terms of the symmetry generators $j^{ab}$ and the fermions $p$ and $\theta$
  of the decoupled system (and their anti-holomorphic analogues)
  has already been presented in \eqref{K}.
  In this case the zero mode sector of the corresponding expressions indeed
  reduces to the tilded differential operators \eqref{tildeRpMore} acting on the
  space $\mc{H}_0$ if we identify the auxiliary fields $p$ and $\bar{p}$ with
  the fermionic derivatives as usual. For the identification to hold it is crucial
  that in the first term $\partial\theta^a$ in $S_1^a$ the zero mode of the
  coordinate field $\theta$ is eliminated by the action of the derivative.
  Needless to say, similar considerations apply for the anti-holomorphic sector.
\smallskip

  After having established the structure of $\mc{\hat{H}}_0$ as a
  representation space with respect to $\widehat{\text{psu}}(1,1|2)_k
  \oplus\widehat{\text{psu}}(1,1|2)_k$
  it is just a small step to spell out the proposal
\begin{equation}
  \label{eq:StateAux}
  \mc{\hat{H}}_0 \ \cong \
\bigoplus_{2J=0}^{k-2}\bigoplus_{w\in\Integer}\int_{\mathbb{S}}dj\int_0^1\!\!\!d\alpha\,
\mc{V}_{[(j,\alpha),J]_w}\otimes\mc{V}_{[(j,\alpha),J]_w^+}
   \oplus\bigoplus_{2J=0}^{k-2}\bigoplus_{w\in\Integer}
  \int_{\frac{1}{2}}^{\frac{k+1}{2}}\!\!\!dj\,\mc{V}_{[(+,-j),J]_w}\otimes\mc{V}_{[(+,-j),J]_w^+}\ \ .
\end{equation}
  This space meets all the requirements
  stated above. First of all, it is indeed a fully covariant version of
  the space \eqref{eq:StateDeco}. This is immediately
  obvious in view of our discussion of affine Lie superalgebra representations
  in section \ref{sc:AffineReps} and, in particular, given the definition
  \eqref{eq:AffKac}. Only the treatment of spectral flow requires a few
  comments since a spectral flow which exclusively acts in the $\AdS_3$
  sector cannot be lifted to the full superalgebra $\PSU$ in general.
  Indeed, as can be inferred from \eqref{eq:SF} the only spectral flow
  automorphisms which solely act on the $\AdS$ factor are of the form
  $\c^{(w,0)}$. But in order to keep the integer moding of the fermions
  which is required for the implementation of the global supersymmetry
  one would have to choose $w$ even, resulting in the ommission of every
  second representation. The simple way out is to consider
  the spectral flows $\c^{(w,w)}$ for all $w\in\Integer$. In this case the
  moding of the fermions stays invariant and the action on the $\SU$ sector
  can be absorbed in a relabeling of the corresponding integrable weights.
  Spectral flow automorphisms that respect the fermionic boundary
  conditions were found in \cite{Schomerus:2005bf} to be exact
  symmetries of the WZNW model on the supergroup $\GL$. We believe
  that this observation generalizes to arbitrary supergroups. In the
  case of $\PSU$ it is indeed consistent with the results of
  Maldacena and Ooguri \cite{Maldacena:2000hw}.
\smallskip

  On the other hand, beside being supersymmetric, the state space
  \eqref{eq:StateAux} can be shown to be a straightforward affinization
  of the minisuperspace result $\mc{H}_0$. In order to establish this
  correspondence we consider the semi-classical limit $k\to\infty$ in
  which the curvature of the supergroup becomes small and the truncation
  of the spectrum can be neglected.
  We are moreover only interested in the light states whose conformal dimension
  approach zero. This forces us to discard all the spectral flow
  representations.\footnote{We should obviously keep those which map $(+,j)$
  to $(-,j')$.} We are thus left with the ground states of the affine modules
  $\mc{V}_{[\mu,\nu]}$ and these obviously transform in the Kac module
  $V_{[\mu,\nu]}$. This concludes our treatment of the decoupled theory.
\smallskip

  Now we turn our attention again to the full WZNW model as defined in
  eq.~\eqref{eq:WZWnew}, including the interaction term $\mc{S}_{\text{int}}$.
  Let us remind the reader that the space $\mc{H}_0$ just has been an
  auxiliary space which helped analyzing the space $\mc{H}=\Xi\mc{H}_0$
  on which the true left and right regular actions of $\psu$ have
  been defined. The same happens in the full $\PSU$ field theory where
  the regular actions are promoted to local symmetries. Roughly
  speaking, the presence of the additional term $\mc{S}_{\text{int}}$
  imitates the action of the operator $\Xi:\mc{H}_0\to\mc{H}$ and
  modifies the definition of the affine currents. This means that
  the true state space of the $\PSU$ WZNW model is given by
  a space $\mc{\hat{H}}$ which differs from $\mc{\hat{H}}_0$ in
  the way the affine currents act.
 Without going into details we symbolically introduce the map
  $\hat{\Xi}:\mc{\hat{H}}_0\to\mc{\hat{H}}$ which intertwines the actions
  in the typical sector. It is important, however, to emphasize
  that the representation content of $\mc{\hat{H}}_0$ and $\mc{\hat{H}}$
  is not isomorphic. In particular, the atypical sector in $\mc{\hat{H}}$
  may not be written as the product of holomorphic and anti-holomorphic
  representations since the zero modes $L_0$ and $\bar{L}_0$ of the energy
  momentum tensors, the affine analogues of the Laplace operator discussed
  in section \ref{sc:MS}, are not diagonalizable. Modular invariance
  then enforces
  that the difference of the nilpotent part vanishes on the state space
  and this is only possible if the representations do not factorize.
  Another consequence of the previous statements is the occurence
  of logarithmic correlation functions in the $\PSU$ WZNW model.
\smallskip

  Let us conclude this section with a brief sketch of the calculation
  of correlation functions. Given any vertex operator $\Phi(z,\bar{z})$
  corresponding to a state in the full Hilbert space $\mc{\hat{H}}$
  -- with or without spectral flow --
  we can find a vertex operator $\Phi_0(z,\bar{z})$ in the decoupled
  theory such that $\Phi(z,\bar{z})=\hat{\Xi}\Phi_0(z,\bar{z})$.
  As in the minisuperspace theory the action of $\hat{\Xi}$ basically
  adds subleading contributions to the full vertex operator. The
  correlation functions are then easily determined using the
  description
\begin{equation}
  \label{eq:CF}
  \bigl\langle\Phi(z_1,\bar{z}_1)\cdots\Phi(z_n,\bar{z}_n)\bigr\rangle_{\PSU_k}
  \ =\ \bigl\langle\Phi_0(z_1,\bar{z}_1)\cdots\Phi_0(z_n,\bar{z}_n)
       \,e^{-\mc{S}_{\text{int}}}\bigr\rangle_{\mc{S}_0}\ \ .
\end{equation}
  In order to make sense out of this expression it is necessary
  to cast the interaction term in a form which may be evaluated
  in the decoupled theory. It is not difficult to convince oneself
  that the corresponding alternative form of the interaction term
  in \eqref{eq:DecoLagrangian} is given by
\begin{equation}
  \mc{S}_{\text{int}}
  \ \sim \ \frac{1}{2\pi k}\int\!\!d^2z\,p_a(z)\,D^{ab}(z,\bar{z})\,\bar{p}_b(\bar{z})\ \ .
\end{equation}
The operators $D^{ab}(z,\bar{z})$ are non-chiral vertex operators
  of the $\AdS_3\times\text{S}^3$ WZNW theory which transforms in the
  $(1/2,1/2)\times(1/2,1/2)$ representation with respect to the
  holomorphic and anti-holomorphic bosonic currents. Given the knowledge
  of correlation functions in the decoupled theory
  \cite{Dotsenko:1990zb,Maldacena:2001km} it is now a tedious but
  algorithmic exercise
  to calculate the right hand side of \eqref{eq:CF}. It is worth
  noticing that due to the presence of the fermions $p$ and $\bar{p}$
  the expansion will terminate after a finite number of terms.
  The only caveat concerns the insertion of the vertex operators
  $D^{ab}$ in the correlation functions of the $\AdS_3$ WZNW
  model since these do not exist in the physical spectrum of the
  bosonic theory but are rather associated with non-normalizable
  degenerate fields. It is well known that correlation functions
  with insertions of such degenerate fields can be determined from
  the physical ones by analytic continuation. In fact, a reversal of
  this argument was a crucial ingredient in the solution of
  the two best understood non-rational conformal field theories,
  i.e.\ Liouville theory \cite{Teschner:1995yf} and
  the (euclidean) $\AdS_3$ model \cite{Teschner:1997ft}.
  Hence, all the ingedients for the computation of correlators in the
  WZNW model on the supergroup are determined by the solution of the
  bosonic model, as we have claimed several times before.

\section{Casimir decomposition of the state space}

The central result of the following section can be considered as a
corollary of our observation that the ground states of the field
theory all transform according to projective representations. As
we shall explain below, this implies that one can count the number
of field theoretic states in any given $\psl$ representation
through some variant of the Racah-Speiser algorithm. The results
play a central role for the study of the RR deformation. Though we
shall publish this investigation in a separate paper, we decided
to include a short discussion of the RR deformation and its
relation to Casimir decompositions at the end of this section.

\subsection{\label{sc:Branching}The $\psl$ symmetry and its branching functions}

By construction, the RR deformation  preserves both global left
and right $\psl$ action. Hence, the state space of the perturbed
theory will continue to carry a representation of these two
commuting $\psl$ transformations. In order to study the
perturbation, it seems worthwhile to decompose the state space of
the model explicitly with respect to the preserved symmetries, in
particular with respect to the left and right $\psl$ action. Since
this is a bit cumbersome to write down for the full state space of
the bulk theory, we shall explain the main idea in a simpler
example that is relevant for the study of instantonic point-like
branes in the $\PSU$ model.
\smallskip

Naively, one might expect that the spectrum of open strings on
such branes contains no zero modes and hence possesses a unique
ground state that transforms in the trivial representation of the
preserved $\psl$ of the boundary theory. If this was true, the
decomposition of the boundary spectrum into representations of
$\psl$ would be extremely difficult, if not impossible. In fact,
states of the boundary theory would then transform according to
all those representations that appear in some tensor power of the
adjoint representation. Our investigations show \cite{Gotz:2005ka}
that very exotic indecomposable representations can emerge in this
way. Since the adjoint representation of $\psl$ is atypical and not
projective, tensor powers are specifically not decomposable
into projectives. The indecomposables that arise in this way
cannot even be listed easily so that the bookkeeping of the
possible states and their transformation laws appears as a
daunting task.
\smallskip

Fortunately, the boundary spectrum of a maximally symmetric
point-like brane in our WZNW model does possess zero modes. In fact,
a thorough investigation of the gluing condition shows that while
such branes are localized in the bosonic coordinates they must
necessarily be delocalized in all fermionic directions (details
will be published elsewhere). Since there are eight fermionic
coordinates, each contributing one zero mode, we conclude that
the ground states transform according to the representation
$$ B(0,0) \ \cong \ \P_0 \ \oplus \ [1,0]\ \oplus\ [0,1] \ \ , $$
i.e.\ as a sum of projective representations. Excited states
therefore transform in representations that emerge from a product
of a projective representation with some power of the adjoint and
which, by abstract mathematical results, can be decomposed into
projectives. This result is once more a confirmation of what we saw
in the bulk theory: the physical states of the $\PSU$ WZNW model all
transform in projective representations, i.e.\ they either form
typical long multiplets or they sit in maximally extended atypical
representations.
\medskip

In the first part of  this work we listed explicitly all the
(finite dimensional) projective representations of $\psl$. Our aim
now is to compute the branching functions for a state space
of the WZNW model into $\psl$ representations. We shall first show
that there is an efficient algorithm that determines this
branching explicitly and then we shall state the results for one
example. Note that the branching functions can be considered
as characters of the so-called Casimir algebra \cite{Bouwknegt:1993wg}.

\subsubsection{The Racah-Speiser algorithm}

In its original form, the Racah-Speiser algorithm is a powerful tool
which allows to decompose tensor products of representations of
semi-simple Lie algebras. The only knowledge required is the
weight content of one of the representations involved and the
action of the Weyl group. In this paper we will use it in a
slightly different setup. We assume that we have given a set of
weights belonging to some finite dimensional representation $R$.
The weight content can be encoded in some generating function,
the character of $R$. In the case of ordinary bosonic Lie algebras
$R$ can be written as a direct sum of irreducible representations $R_i$.
The Racah-Speiser algorithm allows us to determine the $R_i$ and
their multiplicities by just analyzing the original weight system.
Consequently, the character of $R$ can be expressed through the
characters of the irreducible representations $R_i$.
\smallskip

It is clear that for principal reasons the algorithm cannot be
extended to Lie superalgebras. This is due to the presence of not
fully reducible representations: there can exist several
inequivalent representations
which have the same weight content. Our claim, however, is that
the Racah-Speiser algorithm may be extended to Lie superalgebras
as long as one is just dealing with projective representations.
In fact, projective representations share the crucial property
that their characters always contain the fermionic factor $V_F$
(see eq.~\eqref{eq:FermCont})
multiplied by some representations of the bosonic subalgebra.
Hence, the problem of reconstructing the representation content
of a projective representation out of its weight system is reduced
to a problem concerning the bosonic subalgebra. This statement
also allows to calculate tensor products of projective
representations using character methods.
One could thus say that indecomposable projective representations
-- typical irreducibles and projective covers -- play a similar role for
(simple) Lie superalgebras as irreducible ones play for ordinary
(simple) bosonic Lie algebras.
\smallskip

Before we start to discuss complications arising in the super
case it is convenient to explain the idea of the original
Racah-Speiser algorithm in the example of the Lie algebra $\sll$.
Suppose we are given some (finite dimensional) representation
$R$ of $\sll$ with a character of the form
\begin{equation}
  \chi_R(z)\ =\ \sum_{2l\in\Integer}\,a_l\,z^l\ \ .
\end{equation}
  The coefficients $a_l$ give the multiplicity of states
  with isospin $l$. For consistency they have to satisfy $a_l=a_{-l}$.
It is easy to see that we may rewrite the previous expression
  in terms of characters of irreducible representations as
\begin{equation}
  \label{eq:AuxForm}
  \chi_R(z)\ =\ \sum_{j\geq0}\bigl[a_j-a_{j+1}\bigr]\,\chi_j(z)\ \ .
\end{equation}
Due to the linearity of the problem it is enough to prove this
relation on the level of characters of irreducible
representations where it is obvious. Basically the formula counts
the number of weights with isospin $j$ and checks
how many still exist for $j+1$. The latter obviously do not belong
to the spin $j$ representation and have to be subtracted. The
Racah-Speiser trick provides a very simple way e.g.\ to derive the
Casimir characters of $\asu_k$, see \cite{Kac:1990}, after
splitting the algebra into parafermions and a $\hat{u}(1)$ part.
\smallskip

In this paper we are interested in the Casimir characters of $\apsl(2|2)_k$.
As we will see in the following section this problem may be reduced to
one solely involving the bosonic subalgebra $\sll\oplus\sll$.
Applying formula \eqref{eq:AuxForm} to this new situation with two
factors we find
\begin{equation}
  \begin{split}
    \chi_R(z_1,z_2)
    &\ =\ \sum_{2l_1,2l_2\in\Integer}\,a_{l_1l_2}\,z_1^{l_1}\,z_2^{l_2}\\[2mm]
    &\ =\ \sum_{j_1,j_2\geq0}\,\bigl[a_{j_1,j_2}-a_{j_1+1,j_2}-
    a_{j_1,j_2+1}+a_{j_1+1,j_2+1}\bigr]
          \chi_{j_1}(z_1)\,\chi_{j_2}(z_2)\ \ .
  \end{split}
\end{equation}
In the intended application of this formula to affine modules the
multiplicities of the weights are infinite. In that case we let
the coefficients $a_{l_1l_2}$ depend on a formal variable $q$ in
order to be able to distinguish the energy of the states and to
resolve the infinities.
\smallskip

The previous formulas are obvious even without using fancy
technology and recruiting great names. Yet, since our ideas for
the calculation of Casimir characters are
likely to apply for more complicated Lie superalgebras we would
like to make clear from the start that our considerations easily
may be generalized. In that case one has to use the shifted action
of the Weyl group $w\cdot\mu=w(\mu+\rho)-\rho$ in order to map the
given weights into the fundamental domain, taking into account the
sign of the transformation. The result will then be a sum over the
corresponding highest weight vectors, and these in turn can be
replaced by characters of irreducible representations.

\subsubsection{Example: Branching rules for $\P_0$}

As we have announced before, our goal is to decompose the space of
physical states of the $\PSU$ WZNW model with respect to the horizontal
subalgebra. For simplicity we shall focus on one particular
building block of the open string spectrum on a point-like brane,
namely on the decomposition of the representation $\hat{\P}_0$
of the affine $\apsl(2|2)$. The results can easily be
extended to the other two pieces $[0,1]^\wedge$ and $[1,0]^\wedge$
which appear in the brane's spectrum (see above).\footnote{It
is also rather straightforward to find the generalization to
infinite dimensional representations like those that
appear in the bulk spectrum.}
\smallskip

The representation $\aP_0$ is built on top of the finite
dimensional projective cover $\P_0$ by acting with the
negative modes of the current algebra, followed by the removal of
all bosonic singular vectors, i.e.\ all the singular vectors that
do not appear among the ground states. As we observed above the
states at higher energy levels transform in the tensor product of
$\mc{P}_0$ with (symmetrized) tensor products of the adjoint
representation of $\psl$ with itself. Since the tensor
products of arbitrary representations with a projective one are
projective again, the affine representation may be written as
\begin{equation}
  \label{eq:ProjDeco}
  \aP_0\bigr|_{\psl}\ =\ \sum_{j_1\neq j_2}a_{j_1j_2}(q)
  \,[j_1,j_2]+\sum_{j}b_j(q)\,\mc{P}_j\ \ .
\end{equation}
This should be read as a formal decomposition of the affine
representation into representations of the horizontal subalgebra.
The multiplicities on each energy level are contained in the
generating functions $a_{j_1j_2}(q)$ and $b_j(q)$ which can be
considered as characters of the Casimir algebra.
\smallskip

Now we employ our knowledge about the bosonic content of
projective representations as stated in eq.\ \eqref{eq:ProjBos}.
If we denote by $\chi_F$ the character of the fermionic zero modes,
as before, then the character of $\hat{\P}_0$ is given by
\begin{equation}
  \chi_{\aP_0}
  \ =\ \Biggl[\,\sum_{j_1\neq j_2}a_{j_1j_2}(q)\,\chi_{(j_1,j_2)}
       +\sum_{j}b_j(q)\bigl[2\chi_{(j,j)}-\chi_{(j+\foh,j+\foh)}-
       \chi_{(|j-\foh|,|j-\foh|)}\bigr]\Biggr]\,\chi_F\ \ .
\end{equation}
All the relevant notations have been introduced in the first part
of this work. Given the supercharacter on the left hand side we
can then in principle  derive the coefficients $a_{j_1j_2}(q)$ and
$b_j(q)$ using a refined version of the Racah-Speiser algorithm.
From part~I of this work we recall that
\begin{equation}
\chi_{\aP_0}(q,z_1,z_2)
  \ = \ 2 \chi_{[0,0]}(q,z_1,z_2) - 2 \chi_{[1/2,1/2]}(q,z_1,z_2)
  \ \ .
\end{equation}
Explicit formulas for the supercharacters on the right hand side
were provided in eq.~\eqref{chargentyp}. What is most important
for us is that the supercharacter of $\hat{\P}_0$ possesses an overall
factor $\chi_F$ as any projective representation. By expansion
into powers of $z_1$ and $z_2$ and comparison we are thus able to
uniquely determine the functions $a_{j_1j_2}(q)$ and $b_j(q)$. To
this end let us consider the previous expression as a generating
function for the $q$-series $c_{mn}(q)$,
\begin{equation}
  \chi_{\aP_0}(q,z_1,z_2)
  \ =\ \sum_{2m,2n\in\Integer}c_{mn}(q)\,z_1^mz_2^n\,\chi_F(z_1,z_2)\ \ .
\end{equation}
Then the functions $a_{j_1j_2}(q)$ are determined by the
Racah-Speiser trick,
\begin{equation}
  a_{j_1j_2}(q)\ =\ c_{j_1,j_2}(q)-c_{j_1+1,j_2}(q)-
  c_{j_1,j_2+1}(q)+c_{j_1+1,j_2+1}(q)\ \ .
\end{equation}
These quantities also have meaning for $j_1=j_2$. Yet, in that case
they do not count ``real'' representations but just the Kac modules
which sit inside the projective representations. We find
\begin{align}
  a_{00}&\ =\ 2b_0+b_{\foh}&
  a_{\foh\foh}&\ =\ 2b_0+2b_\foh+b_1&
  a_{jj}&\ =\ b_{j-\foh}+2b_j+b_{j+\foh}\ \ .
\end{align}
This relation needs to be inverted to find the values of the
$b_j$. For a fixed energy level this inversion is almost trivial,
one just has to start from the contributions with highest spin in
order to find the corresponding projective covers. The described
procedure may seem a bit abstract, but it is straightforward to
implement the expansions and the Racah-Speiser trick on the
computer. In this way the branching functions $a(q)$ and $b(q)$
can in principle be determined to any desired order. For large
values of the level $k$ one finds for instance
\begin{equation}
  \begin{split}
    \hat{\mc{P}}_0(q)
    &\ =\ \mc{P}_0\ \oplus\ %
          q\Bigl[4\,\mc{P}_{\frac{1}{2}}\oplus6\bigl([1,0]\oplus[0,1]\bigr)
 \oplus2\bigl((\fth,\foh)\oplus(\foh,\fth)\bigr)\Bigr]\\[2mm]
    &\qquad\ \oplus\ q^2\Bigl[\mc{P}_0\oplus16\,\mc{P}_{\frac{1}{2}}\oplus4\,
\mc{P}_1\oplus4\bigl([2,1]\oplus[1,2]\bigr)\oplus6\bigl([2,0]\oplus[0,2]\bigr)\\[2mm]
    & \quad\oplus24\bigl([1,0]\oplus[0,1]\bigr)\oplus2\bigl((\ffh,\foh)
\oplus(\foh,\ffh)\bigr)\oplus18\bigl((\fth,\foh)
\oplus(\foh,\fth)\bigr)\Bigr]\oplus\cdots\ \ .
  \end{split}
\end{equation}
With a bit more work one might also be able to write down
closed formulas.
\smallskip

Before we conclude this subsection, let us stress once more that
the whole procedure relied extremely on the fact that only
projective representations occurred in the affine character. If
there had been non-projective atypicals in addition to the
projective ones, knowledge of the characters would have been
insufficient to determine the decomposition.

\subsection{Some remarks on the RR deformation}
\def\s{\sigma}

\begin{figure}
  \begin{center}
    \input{flux_diagram.pstex_t}
    \caption{\label{fig:fluxes}The moduli space of string theory on $\PSU$.
      The vertical axis gives the normalization of the kinetic term, the
      horizontal the normalization of the Wess-Zumino term. The lines with
      $\lambda\neq0$ correspond to marginal deformations of the WZNW model.}
  \end{center}
\end{figure}
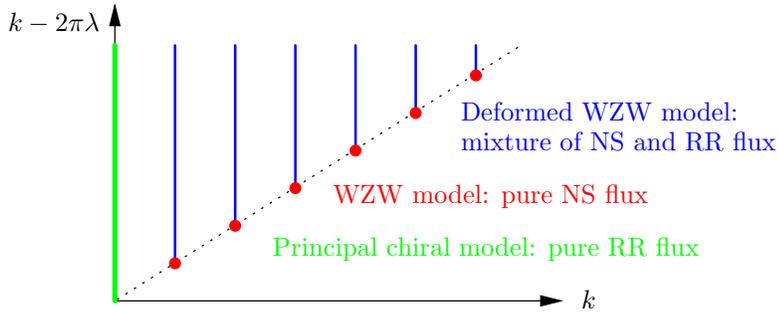

The RR deformation of the $\AdS_3$ background corresponds to
adding the the following extra to the action of the WZNW model
$$ \mc{S}_{RR;\lambda}^{\PSL} \ = \ - \frac{\lambda}{2\pi}
\, \int d^2z \ \tr \left(S^{-1} \partial S S^{-1}
   \bar \partial S\right) \ \ . $$
We can rewrite the perturbation in terms of the fields we have
studied above. To this end, we shall need the left and right
invariant (anti-)holomorphic currents $J^\mu(z)$ and $\bar
J^\nu(\bar z)$ along with some degenerate primary fields
$\Phi_{\mu\nu}(z,\bar z)$ that transform in the (atypical) adjoint
representation $[1/2]$ of $\psl$, i.e.
\begin{equation} \label{JPOPE}
  \begin{split}
J^\mu(z)\,  \phi_{\nu\rho}(w,\bar w) &\ =\
\frac{if^{\mu\nu\s}}{z-w}
 \ \phi_{\s\rho}(w,\bar w) + \dots  \ \ \ , \\[2mm]
\bar J^\mu(\bar z) \, \phi_{\nu\rho}(w,\bar w) &\ =\
\frac{if^{\mu\rho\s}}
 {\bar{z}-\bar{w}} \ \phi_{\nu\s}(w,\bar w) + \dots\ \ .
  \end{split}
\end{equation}
It is then easy to identify the RR perturbation with the one that
is generated by the composite field
\begin{equation}
\Phi(z,\bar z) \ = \  :J^\mu(z) \,
            \phi_{\mu\nu}(z,\bar z)\,  \bar J^\nu(\bar z): \ \ .
\end{equation}
By construction, the field $\Phi$ has conformal weights $h = \bar
h = 1$ in the WZNW model but in principle its dimension could
change when we perturb the theory. According to
\cite{Bershadsky:1999hk} (see also \cite{Gerganov:2000mt,
Ludwig:2002fu} for related studies), however, $\Phi$ is truly
marginal, i.e.\ its dimension remains at $h = \bar h = 1$ in all
orders of perturbation theory. Note that the perturbation with the
field $\Phi$ rescales the kinetic term of the WZNW model and thus
alters the relative normalization of kinetic and Wess-Zumino term.
The resulting moduli space and its physical interpretation is
summarized in figure \ref{fig:fluxes}.
\smallskip

We do not want to re-derive the conformal invariance of the RR
deformation here, but instead will discuss a somewhat more general
assertion. To begin with, it follows from the marginality of
$\Phi$ that the deformed theory possesses the usual
(anti-)holomorphic Virasoro field. The latter is $\psl$ invariant
and hence acts as a symmetry within the branching spaces of the
decomposition \eqref{eq:ProjDeco}. The true chiral symmetry of the
deformed theory is larger. In fact, it was claimed in
\cite{Bershadsky:1999hk} that all the Casimir fields in the
current algebra provide chiral fields of the deformed model, i.e.\
that all fields of the form
\begin{equation}
  W^{(n)} \ = \ t_{\mu_1\cdots\mu_n}\,J^{\mu_1}\cdots J^{\mu_n}
\end{equation}
involving contractions with an invariant, symmetric and traceless
tensor $t$ are holomorphic even after a finite deformation with
$\Phi$. We would like to establish this in first order
perturbation theory. Let us emphasize that the symmetry and
tracelessness of $t$ imply that on the right hand side no
particular normal ordering prescription has to be specified.
\smallskip

To begin with, we analyze the behaviour of a single current. An
insertion of the latter into a correlation function corresponds to
the perturbative expansion
\begin{equation}
  \bigl\langle J^\mu(z,\bar{z})\cdots\bigr\rangle_\lambda
  \ =\ \bigl\langle J^\mu(z)\cdots\bigr\rangle_0
       -\lambda\,\Bigl\langle J^\mu(z)\cdots\int d^2w\,\Phi(w,\bar{w})\Bigr\rangle_0+\cdots\ \ .
\end{equation}
  The integral may easily be evaluated with the help
  of the operator product expansion
\begin{equation}
  J^\mu(z)\,\Phi(w,\bar{w})
  \ =\ \frac{k:{\phi^\mu}_\nu\bar{J}^\nu:(w,\bar{w})}{(z-w)^2}
  \ =\ \frac{k:{\phi^\mu}_\nu\bar{J}^\nu:(z,\bar{z})}{(z-w)^2}-\frac{k:\partial{\phi^\mu}_\nu\bar{J}^\nu:(z,\bar{z})}{z-w}\ \ .
\end{equation}
  The second expansion where the argument
  $w$ is replaced by $z$ is more suitable for the
  calculation of the integral. Introducing the usual
  step function cut-off which restricts the
  integration to the domain $|z-w|^2>a^2$ we find,
  following an argument of Cardy \cite{Cardy:1989da},
\begin{equation}
  \label{eq:JDer}
  \bartial J^\mu
  \ =\ \pi\lambda\,k\,:\partial{\phi^\mu}_\nu\bar{J}^\nu:+\cdots
  \ =\ -2\pi i\lambda\,k\,{f^\mu}_{\rho\sigma}:J^\rho{\phi^\sigma}_\nu\bar{J}^\nu:+\cdots\ \ .
\end{equation}
  This equation is the quantum analogue of the
  classical relation $\bartial J=c(k,\lambda)[J,g\bar{J}g^{-1}]$
  which can be derived from the equations of motion of the perturbed
  Lagrangian in connection with the Maurer-Cartan equation. We
  believe that the higher order terms in $\lambda$ reproduce
  precisely the classical equation, i.e.\ that $\bartial J^\mu$
  is proportional to
  ${f^\mu}_{\rho\sigma}:J^\rho{\phi^\sigma}_\nu\bar{J}^\nu:$
  and that the $\lambda$-dependent prefactor coincides with
  the classical expression $c(k,\lambda)$.
\smallskip

  We are now finally prepared to investigate the chiral properties
  of the Casimir currents $W^{(n)}$. Taking the derivative
  will produce a number of terms but we can move
  the current with the derivative to the last position. This is because
  the difference between two different normal orderings of
  the currents and their derivatives always involves the
  metric or the structure constants with two open indices.
  Upon contraction with $t$ these terms vanish by assumption.
  Together with the result \eqref{eq:JDer} this implies
\begin{equation}
  \bartial W^{(n)}
  \ =\ -2\pi i\lambda\,k\,{f^{\mu_n}}_{\rho\sigma}\,t_{\mu_1\cdots\mu_n}\,J^{\mu_1}\cdots J^{\mu_{n-1}}J^{\rho}{\phi^\sigma}_\nu\bar{J}^\nu\ \ .
\end{equation}
  In this equation the normal ordering again is not relevant.
  This is basically due to the same reasons as above, combined
  with the fact that operator products of currents with the
  field $\phi$ give rise to a single term involving the
  structure constants $f$. In addition we have to use the vanishing
  of the dual Coxeter number of $\psl$ in order to get
  rid of multiple $f$-contractions. Reordering and relabeling
  the currents we then find the desired result $\bartial W^{(n)}=0$
  due to the invariance of the tensor $t$.
\smallskip

Having established the holomorphicity of Casimir fields, it seems
important to add a few comments on denominations. It would be
tempting to baptize the chiral algebra that is generated by the
Casimir fields as Casimir algebra. But this is not how the latter
term is used. By definition, the $\psl$ Casimir algebra is the
chiral algebra whose characters are given by the branching
functions we computed in the previous subsection
\cite{Bais:1988dc,Bais:1988zk}. These are not the same as the
characters of the chiral algebra that is generated by the Casimir
fields \cite{Bouwknegt:1993wg}. In fact, characters of the latter
may be calculated using the quantum Drinfeld-Sokolov reduction
\cite{Feigin:1990pn,Frenkel:1992ju} and they only appear as
building blocks for the branching functions above, but in most
cases the precise relation is highly involved. Put differently,
the chiral algebra generated by the Casimir fields is much smaller
than the Casimir algebra. Roughly, the difference is that the
Casimir fields are obtained as invariant combinations of currents
whereas the Casimir algebra also contains invariant fields
involving derivatives of currents. The first field in the Casimir
algebra that is not a Casimir field appears at conformal weight
$h=4$ and it is given by $\str(\partial J \partial J)$. In this
sense, the decompositions discussed in the previous subsection
would not seem sufficient in order to control the deformation away
from the WZNW point. On the other hand, the spectra possess a
higher degree of degeneracy which cannot be explained by the
chiral symmetries we have described. Take, for example, the
current itself: it does not stay holomorphic beyond the WZ point,
yet its conformal weight rests at $h=1$ all along the line of
marginal deformations. Such additional degeneracies may possibly
be attributed to some new symmetry whose algebraic structure
remains to be uncovered.\footnote{Additional Yangian symmetries
(see \cite{Bernard:1992ya,MacKay:2004tc} for a review) are known
to exist in these models, see e.g.\ \cite{Bena:2003wd}. See also
\cite{Dolan:2003uh,
Vallilo:2003nx,Hatsuda:2004it,Das:2005hp,Kagan:2005wt,Miller:2006bu,
Bianchi:2006im}
for some closely related studies and results.} In a forthcoming
paper, we shall suggest a different picture that is intimately
related to Casimir algebras.

\section{Outlook and conclusions}

  In this work we presented a complete solution of the WZNW model
  on the supergroup $\PSU$ in terms of a free fermion construction
  which kept the bosonic symmetry manifest.
  After a thorough discussion of the representation theoretic
  foundations in part~I we derived the precise form of the spectrum
  based on methods of harmonic analysis. It was found that the state
  space splits into two parts. The typical (non-BPS) sector behaves nicely and
  decomposes into tensor products of chiral irreducible representations.
  On the other hand, there exists an atypical (BPS) sector related to
  states with vanishing conformal dimension where our intuition
  from purely bosonic WZNW models breaks down. Here we found non-chiral
  indecomposable representations on which the zero-mode of the energy
  momentum tensor is not diagonalizable.
  Although we just provided a brief sketch of how correlation functions
  may be calculated it is thus established that the $\PSU$ WZNW model
  is a logarithmic conformal field theory.
\smallskip

We would like to stress that the logarithms only arise because the
WZNW model on the supergroup $\PSU$ {\it does not} factorize into
a product of fermionic and bosonic subsectors, contrary to a
widespread claim. In the minisuperspace theory, the coupling may
be ascribed to the existence of the nilpotent term $Q$ in the
Laplacian. The operator $Q$ is a differential operator in the
fermionic directions with coefficients which vary along the
bosonic base. Compared to the factorized theory, this additional
term alters the structure of the eigenfunctions along with their
transformation law under $\psl$. As a result, all the states
transform in projective representations only, both in the particle
limit and the full field theory. It is worthwhile to emphasize
that without the coupling the field theory ground states would
have transformed as (a product of) Kac modules, even in the
atypical sector where the modules degenerate. Hence,
the interaction between bosons and
fermions drastically changes the embedding structure of the fermionic
singular vectors in our state space.
\smallskip

As in usual free field constructions for purely bosonic WZNW models
one might be tempted to set all singular vectors to zero and to work
with irreducible representations only. But in the case of supergroup
WZNW models, the fermionic singular vectors cannot be decoupled,
at least for generic values of the level $k$. An investigation of
the relevant Knizhnik-Zamolodchikov equations similar to the one
in \cite{Rozansky:1992rx} shows indeed that generically local
solutions contain logarithms. Hence, at least some of the
fermionic singular vectors are needed for consistency. There are
also other ways to argue that fermionic singular vectors have a very
different status from their bosonic counterparts. In particular,
it is {\it unavoidable} that the states on higher energy levels in
affine modules -- irreducible or not -- transform in reducible but
indecomposable representations of the horizontal superalgebra.
Therefore it seems unnatural to insist on removing fermionic
singular vectors among the ground states and to work with
irreducible affine representations only. Even worse, in atypical
irreducible affine modules or the associated Kac modules we have
little or no chance to ever control the behavior of all excited
states under global $\psu$ transformations.

It is in this context
that projectivity of representations comes to our rescue. Indeed,
for affine modules based on a projective representation of
$\psu\oplus\psu$ all the excited states transform in projective
representations. Moreover, the Racah-Speiser algorithm allows to
determine the decomposition rather explicitly. Since the $\psu$
symmetry is an important part of the symmetry that remains
unbroken when we turn on RR flux, the occurrence of projective
representations is the best we could hope for if we are interested
in getting a handle on the $\sigma$-model describing
$\AdS_3\times\text{S}^3$ with a mixture of NS and RR fluxes.
Hence, we think that even in cases where it might not be strictly
necessary for reasons of consistency it is much preferable to
define the WZNW model such that it includes all the fermionic
singular vectors. This is also suggested by the minisuperspace
analysis in which fermionic singular vectors appear naturally
among the eigenstates of the Laplacian.

Another remarkable consequence of the coupling of bosons and
fermions is that the naive algebra of functions on $\PSU$ --
generated by the product of functions on its body $\AdS_3\times
S^3$ with monomials in the fermionic variables -- does not furnish
the appropriate model for the representation space of the global
supersymmetry transformations. The origin of this surprising fact
is that the fermions transform in a finite dimensional {\em
non-unitary} representation of the bosonic subgroup. In fact,
along with the two supersymmetry transformation laws -- one for
the decoupled theory and one for the full theory -- we have to
distinguish two different models for the representation spaces.
While in the decoupled theory we are indeed working with the naive
algebra of functions, this is not true anymore in the full,
coupled system. The operator $\Xi$ derived from $Q$ (or
analogously the interaction term\footnote{One could also refer to
it as screening charge but it should be kept in mind that no BRST
procedure is implied in our description.} in the full WZW model)
mediates between these two inequivalent representation spaces. As
part of this process the operator $\Xi$ entangles the two chiral
subsectors of the free fermion theory by adding subleading
contributions to the free field vertex operators. These in turn
change the normalizability properties with respect to the bosonic
subgroup and explain why the naive algebra of functions did not
provide the proper representation space for the full theory.
\smallskip

Though the observations we have listed in the last few paragraphs
emerged from the study of the WZNW model on $\PSU$, it is clear
that they extend to a much larger class of models. Let us point
out that the minisuperspace analysis is not at all restricted to
the WZ-point and hence most of our remarks concerning the
structure and importance of fermionic singular vectors apply to
more general sigma models, in particular to principal chiral
models on a large class of supergroups. Similarly, subtleties such
as the coupling of bosonic and fermionic degrees of freedom --
which eventually lead to the occurrence of indecomposable
non-chiral representations and logarithmic correlation functions
-- are bound to arise in more general setups. In fact, the same
features are common to most relativistic theories with a global
target space superalgebra symmetry.
\smallskip

Supergroups and cosets thereof appear naturally in all attempts to
quantize super-string theory in a manifestly target space
supersymmetric way. On the corresponding backgrounds, the
supersymmetry transformations are realized geometrically as an
isometry or, more precisely, as the left and (for the group case)
right action of the supergroup on itself. This statement holds in
particular for all supersymmetric $\AdS$-spaces which can be
expressed as (right) cosets based on superconformal groups such as
$\text{PSU}(N,N|2N)$ \cite{Metsaev:1998it,Berkovits:1999zq},
higher dimensional relatives of the supergroup considered here.
Due to the presence of Ramond-Ramond fluxes the quantization of
these backgrounds has been a notoriously difficult task.
\smallskip

In the case of $AdS_3$ backgrounds, the hybrid approach of
\cite{Berkovits:1999im} provides one way to resolve the conceptual
issues. As we stressed several times before, it involves the sigma
model on $\PSU$. Concerning higher dimensional backgrounds, the
quantization of string backgrounds using pure spinors
\cite{Berkovits:2000fe} may be considered the most promising
recent development, see also
\cite{Vallilo:2002mh,Berkovits:2004xu} for some results on
$\AdS_5$ backgrounds that were obtained in this formalism. Yet,
one drawback of the original formulation was the necessity of
solving the pure spinor constraint explicitly using a suitable
choice of coordinates. This in turn partly ruined the manifest
Lorentz covariance. The problem was overcome in
\cite{Grassi:2001ug} upon introduction of new additional ghost
systems. In follow up papers the central role of a special affine
Lie superalgebra has been emphasized
\cite{Grassi:2003kq,Grassi:2004cz} (see also
\cite{Guttenberg:2004ht}). The existence of local and global
superalgebra symmetries connects these developments with the
technical aspects of our work, even though we have not been
concerned with imposing the physical state conditions.

\subsubsection*{Acknowledgements}

  We would like to thank Giuseppe d'Appollonio, Denis Bernard,
  Matthias Gaberdiel,
  J\'er\^ome Germoni, Niall MacKay, Yaron Oz, Soo-Jong Rey,
  Sylvain Ribault, Hubert
  Saleur, Kareljan Schoutens, Didina Serban, Vera Serganova,
  Paul Sorba, Anne Taormina, J\"org Teschner,
  Alexei Tsvelik, G\'erard Watts, Kay Wiese and Charles Young for useful
  and inspiring discussions during various stages of this project.
  We are also grateful for the
  hospitality at the ESI during the workshop ``String theory in curved
  backgrounds'' which stimulated the present work. Moreover, Thomas
  Quella acknowledges the kind hospitality of the SPhT in Saclay and at
  the DESY in Hamburg during numerous visits.
\smallskip

  This work was partially supported by the EU Research Training Network grants
  ``Euclid'', contract number HPRN-CT-2002-00325, ``Superstring Theory",
  contract number MRTN-CT-2004-512194, and ``ForcesUniverse'', contract
  number MRTN-CT-2004-005104. Until September 2006 Thomas Quella has
  been funded by a PPARC postdoctoral fellowship under reference
  PPA/P/S/2002/00370. He also received partial support by the PPARC
  rolling grant PP/C507145/1.

\newpage
\appendix
\section{\label{sc:KacKazhdan}On the irreducibility of generalized Fock spaces}

  The Kac-Kazhdan formula encodes the precise structure of singular vectors
  in a Verma module, including their multiplicities \cite{0427.17011,Kac:MR1026476}.
  This in turns allows
  one to represent the characters of irreducible representations as
  alternating sums of characters of Verma modules \cite{Kac:1984mq}.
  In this section we are going to discuss the singular vectors of Verma
  modules over $\apsl(2|2)$ and show that
      their irreducible quotients are typically
  isomorphic to the generalized Fock modules introduced in section
  \ref{sc:AffineReps}. Instead of working directly with the Kac-Kazhdan
  determinant,
  we are taking a more direct and physically more intuitive route here,
  which to the best of our knowledge should be completely equivalent.
  Nevertheless we have to introduce a bit of notation first.
\smallskip

\begin{figure}
  \centerline{\input{psl_roots.pstex_t}}
  \caption{\label{fig:roots}The root diagram of $\psl$.}
\end{figure}
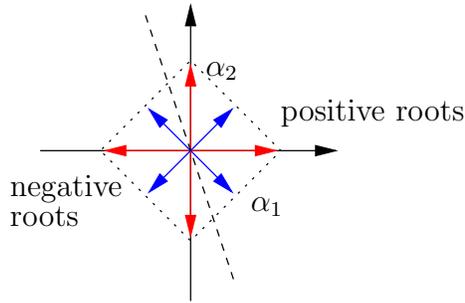

  The set of all pairs $(\mu_1,\mu_2)$ with $\mu_i\in\Integer$ forms the weight
  lattice of $\psl$. The two entries correspond to weights
  in the individual factors of $\sll\oplus\sll$, respectively
  (let us recall that the weight is twice the spin).
  Due to the embedding into the supersymmetric setup
  we have the slightly unusual scalar product
\begin{equation}
  \bigl\langle(\lambda_1,\lambda_2),(\mu_1,\mu_2)\bigr\rangle
  \ =\ -\frac{1}{2}\,\bigl(\lambda_1\mu_1-\lambda_2\mu_2\bigr)\ \ .
\end{equation}
  In order to describe the root system of the Lie superalgebra
  $\psl$ at least one of the two simple roots has to be chosen
  fermionic. It turns out to be useful to work with a root system
  whose simple roots correspond to $\alpha_1=(1,-1)$ and $\alpha_2=(0,2)$.
  The remaining positive roots are then given by $\alpha_1+\alpha_2=(1,1)$
  and $2\alpha_1+\alpha_2=(2,0)$. The fermionic roots have multiplicity
  two while the bosonic ones just have multiplicity one. A sketch
  of the root diagram can be found in figure~\ref{fig:roots}.
\smallskip

  If we denote the Weyl vector as $\rho=(1,1)$ as usual then the
  conformal dimension of a highest weight representation $\mu$ is
  given by
\begin{equation}
  \label{eq:ConfWeight}
  h_\mu\ =\ \frac{\langle\mu,\mu+2\rho\rangle}{2k}\ \ .
\end{equation}
  Note that the conformal dimension is invariant under the
  transformation
\begin{equation}
  w\ast\mu\ =\ w(\mu+\rho)-\rho\ \ ,
\end{equation}
  where $w$ refers to an element of the bosonic Weyl group,
  i.e.\ a pair of elements of the Weyl group of $\sll$
  (the statement holds in general though). This may be traced
  back to the fact that the corresponding Weyl transformations leave
  the metric $\langle\cdot,\cdot\rangle$ invariant. Reflections
  perpendicular to the fermionic roots, however, change the
  sign. Therefore they should not be used in the formula above.
\smallskip

  After these remarks we are finally prepared to discuss the
  structure of Verma modules over $\apsl(2|2)$. We
  will consider a Verma module that is based on a highest weight $(\mu,h_\mu)$
  where we included the eigenvalue of $L_0$, the conformal
  weight, for completeness. Singular vectors $(\nu,h_\nu)$
  can just occur if the difference $(\mu-\nu,h_\mu-h_\nu)$
  is a linear combination of affine simple roots with non-negative
  coefficients. This in particular implies that the conformal
  weights have to satisfy the relation
\begin{equation}
  h_\nu\ =\ h_\mu+n
\end{equation}
  with a non-negative integer $n$. Note that the conformal
  dimension of a singular vector indeed is fixed to be $h_\nu$
  as in \eqref{eq:ConfWeight}
  for algebraic reasons and cannot be chosen arbitrarily.
  Moreover, the difference $\mu-\nu$ has to be an element
  of the root lattice of $\psl$ (in fact the actual
  condition is more restrictive).
  In the following we will assume that every weight which
  according to the previous criteria has the potential to
  describe a singular vector in fact is singular. This
  seems to be a straightforward consequence of the Kac-Kazhdan
  formula \cite{0427.17011,Kac:MR1026476}.
  It is even enough to restrict the analysis to the case
  where the affine weights differ by a multiple of a simple
  root. The other states which decouple are just descendents
  of the ones obtained through the latter.
\smallskip

  To illustrate the decoupling conditions we have to specify
  the affine simple roots first. The bosonic simple roots are
  given by the set
\begin{equation}
  \bigl\{(\alpha,-n)\,\bigl|\,\alpha\in\Delta^{(0)}\,,\,n>0\bigr\}
  \cup\bigl\{(\alpha_2,0)\bigr\}\ \ .
\end{equation}
  The second label in each tupel refers to the energy of the
  roots, i.e.\ to the mode number. The remaining simple roots
  are fermionic,
\begin{equation}
  \bigl\{(\alpha,-n)\,\bigl|\,\alpha\in\Delta^{(1)}\,,\,n>0\bigr\}
  \cup\bigl\{(\alpha_1,0)\bigr\}\ \ .
\end{equation}
  In the previous definitions $\Delta^{(0)}$ and $\Delta^{(1)}$
  refer to the bosonic and fermionic roots of $\psl$.
  $\alpha_1$ and $\alpha_2$ have been specified above. Let us
  stress that they do not coincide with the simple roots of
  $\sll\oplus\sll$.
\smallskip

  Let us discuss the bosonic decoupling conditions first.
  For the $m$-fold application of the root $\bigl((\pm2,0),n\bigr)$
  the decoupling condition reads
\begin{align}
  h_{(\mu_1\pm2m,\mu_2)}&\ \overset{!}{=}\ h_{(\mu_1,\mu_2)}+mn&
  &\Rightarrow&
  \mp(\mu_1+1)\ =\ nk+m\ \ .
\end{align}
  This equation cannot be solved for $m$ (for positive level $k$
  and due to the restrictions on $\mu_1$),
  thus proving the absence of bosonic singular vectors with
  respect to the first factor $\asl_{-k}$.
  On the other hand the $m$-fold application of the root
  $\bigl((0,\pm2),n\bigr)$ yields
\begin{align}
  \label{eq:BosDecoup}
  h_{(\mu_1,\mu_2\pm2m)}&\ \overset{!}{=}\ h_{(\mu_1,\mu_2)}+mn&
  &\Rightarrow&
  \pm(\mu_2+1)\ =\ nk-m\ \ .
\end{align}
  In this case the equation may always be solved for $m$ (for
  positive level $k$). Consequently, {\em all} Verma modules
  of $\apsl(2|2)_k$ possess bosonic singular vectors.
\smallskip

  The situation is different for the fermionic simple roots
  because they just may be applied once, i.e.\ $m$ is bound
  to be one. The corresponding four decoupling conditions are
\begin{subequations}
\begin{align}\label{eq:FermDecoupOne}
  h_{(\mu_1+1,\mu_2+1)}&\ \overset{!}{=}\ h_{(\mu_1,\mu_2)}+n&
  &\Rightarrow&
  \mu_1-\mu_2\ =\ -2nk\\[2mm]
  h_{(\mu_1+1,\mu_2-1)}&\ \overset{!}{=}\ h_{(\mu_1,\mu_2)}+n&
  &\Rightarrow&
  \mu_1+\mu_2+2\ =\ -2nk\\[2mm]
  h_{(\mu_1-1,\mu_2+1)}&\ \overset{!}{=}\ h_{(\mu_1,\mu_2)}+n&
  &\Rightarrow&
  \mu_1+\mu_2+2\ =\ 2nk\\[2mm]
  h_{(\mu_1-1,\mu_2-1)}&\ \overset{!}{=}\ h_{(\mu_1,\mu_2)}+n&
  &\Rightarrow&
  \mu_1-\mu_2\ =\ 2nk\ \ .\label{eq:FermDecoupFour}
\end{align}
\end{subequations}
  In this case none of these equations necessarily possesses
  a solution. We thus realize that the existence of fermionic
  singular vectors is a rather special incidence, related
  to the factual absence of the variable $m$. It is thus
  sensible to introduce the
  notion of a typical Verma module. This is a Verma module
  in which none of the fermionic vectors decouples. In other
  words: The highest weight has to violate {\em all} the conditions
  \eqref{eq:FermDecoupOne}-\eqref{eq:FermDecoupFour}.
\smallskip

  The analysis above has to be slightly refined for $n=0$.
  The reason is that for $n=0$ we are bound to use the positive
  roots of $\psl$ in the equations above but the
  negative ones have to be discarded. Thus just half of the equations
  above will correspond to a valid decoupling condition
  under these circumstances.
\smallskip

  After the rather formal discussion of the previous paragraphs
  we are now prepared to prove the first important mathematical
  result.
\begin{lemma}
  Let $\mu$ be the highest weight of a typical Verma module.
  Then every singular vector $\nu$ in this Verma module is
  again typical.
\begin{proof}
  Since singular vectors can just occur in the second factor
  of $\asl_{-k}\oplus\asl_k$ we just
  have to distinguish two cases, corresponding to the two
  different signs in \eqref{eq:BosDecoup}. In the ``$-$''-case
  one finds
\begin{equation}
  (\nu_1,\nu_2)\ =\ \bigl(\mu_1,-\mu_2+2(nk-1)\bigr)\ \ .
\end{equation}
  The conditions for the existence of a singular vector
  in the submodule $\nu$ on the other hand read
\begin{align}
  -\nu_1+\nu_2\ =\ -\mu_1-\mu_2-2+2nk&\ \overset{?}{=}\ 2lk\\[2mm]
  -\nu_1-\nu_2-2\ =\ -\mu_1+\mu_2-2nk&\ \overset{?}{=}\ 2lk\ \ .
\end{align}
  Each of them could only be satisfied if $\mu$ was atypical.
  Similar considerations apply to the ``$+$''-case.
\end{proof}
\end{lemma}
  Basically the previous Lemma implies that for typical
  modules the complete structure of singular vectors is
  captured by the bosonic singular vectors. It should
  moreover be noted that for $\apsl(2|2)_k$ the decoupling
  conditions for the bosonic roots are precisely
  those that one obtains in Verma modules over
  $\asl_{-k-2}\oplus\asl_{k-2}$.
  As a result we have the
\begin{conjecture}
  The characters of the generalized Fock module based on
  a typical weight $\mu$ and that of the corresponding
  irreducible module obtained from the Verma module coincide.
  In particular the Fock module is irreducible itself.
\end{conjecture}
  In order to promote this conjecture to a theorem one would
  have to discuss the multiplicities of zeroes in the
  Kac-Kazhdan determinant \cite{0427.17011,Kac:MR1026476}
  but we refrain from doing so here.
  For atypical modules one has to work a bit harder to
  obtain the characters of irreducible modules, see section
  \ref{sc:Representations} for details. We conclude by
  expressing our expectation that the reasoning of this appendix
  generalizes to more general classes of affine Lie superalgebras.

\def\cprime{$'$} \def\cprime{$'$}
\providecommand{\href}[2]{#2}\begingroup\raggedright\endgroup


\end{document}

%% file: flux_diagram.pstex_t
\begin{picture}(0,0)%
\includegraphics{flux_diagram.pstex}%
\end{picture}%
\setlength{\unitlength}{4144sp}%
\begingroup\makeatletter\ifx\SetFigFont\undefined%
\gdef\SetFigFont#1#2#3#4#5{%
  \reset@font\fontsize{#1}{#2pt}%
  \fontfamily{#3}\fontseries{#4}\fontshape{#5}%
  \selectfont}%
\fi\endgroup%
\begin{picture}(3928,1912)(796,-2411)
\put(2971,-1231){\makebox(0,0)[lb]{\smash{{\SetFigFont{10}{12.0}{\familydefault}{\mddefault}{\updefault}{\color[rgb]{0,0,1}Deformed WZW model:}%
}}}}
\put(2971,-1411){\makebox(0,0)[lb]{\smash{{\SetFigFont{10}{12.0}{\familydefault}{\mddefault}{\updefault}{\color[rgb]{0,0,1}mixture of NS and RR flux}%
}}}}
\put(811,-691){\makebox(0,0)[rb]{\smash{{\SetFigFont{10}{12.0}{\familydefault}{\mddefault}{\updefault}{\color[rgb]{0,0,0}$k-2\pi\lambda$}%
}}}}
\put(2206,-1726){\makebox(0,0)[lb]{\smash{{\SetFigFont{10}{12.0}{\familydefault}{\mddefault}{\updefault}{\color[rgb]{1,0,0}WZW model: pure NS flux}%
}}}}
\put(1846,-2041){\makebox(0,0)[lb]{\smash{{\SetFigFont{10}{12.0}{\familydefault}{\mddefault}{\updefault}{\color[rgb]{0,1,0}Principal chiral model: pure RR flux}%
}}}}
\put(3691,-2356){\makebox(0,0)[lb]{\smash{{\SetFigFont{10}{12.0}{\familydefault}{\mddefault}{\updefault}{\color[rgb]{0,0,0}$k$}%
}}}}
\end{picture}%

%% file: psl_roots.pstex_t
\begin{picture}(0,0)%
\includegraphics{psl_roots.pstex}%
\end{picture}%
\setlength{\unitlength}{4144sp}%
\begingroup\makeatletter\ifx\SetFigFont\undefined%
\gdef\SetFigFont#1#2#3#4#5{%
  \reset@font\fontsize{#1}{#2pt}%
  \fontfamily{#3}\fontseries{#4}\fontshape{#5}%
  \selectfont}%
\fi\endgroup%
\begin{picture}(2788,1824)(706,-1873)
\put(2161,-1321){\makebox(0,0)[lb]{\smash{{\SetFigFont{12}{14.4}{\rmdefault}{\mddefault}{\updefault}{\color[rgb]{0,0,0}$\alpha_1$}%
}}}}
\put(1891,-511){\makebox(0,0)[lb]{\smash{{\SetFigFont{12}{14.4}{\rmdefault}{\mddefault}{\updefault}{\color[rgb]{0,0,0}$\alpha_2$}%
}}}}
\put(721,-1411){\makebox(0,0)[lb]{\smash{{\SetFigFont{12}{14.4}{\rmdefault}{\mddefault}{\updefault}{\color[rgb]{0,0,0}roots}%
}}}}
\put(721,-1231){\makebox(0,0)[lb]{\smash{{\SetFigFont{12}{14.4}{\rmdefault}{\mddefault}{\updefault}{\color[rgb]{0,0,0}negative}%
}}}}
\put(2341,-781){\makebox(0,0)[lb]{\smash{{\SetFigFont{12}{14.4}{\rmdefault}{\mddefault}{\updefault}{\color[rgb]{0,0,0}positive roots}%
}}}}
\end{picture}%